\documentclass[prb,aps,twocolumn,superscriptaddress]{revtex4-2}

\usepackage[T1]{fontenc}
\usepackage[latin9]{inputenc}
\usepackage{amssymb}
\usepackage{graphicx}
\usepackage{amsmath,color}
\usepackage{mathrsfs}
\usepackage{float}
\usepackage{indentfirst}
\usepackage{subfigure}
\usepackage{multirow}
\usepackage{tabu}
\usepackage{threeparttable}
\usepackage{booktabs}
\usepackage{txfonts}
\usepackage{amsmath,amssymb,bbm}
\usepackage[normalem]{ulem}
\usepackage[euler]{textgreek}
\usepackage{bm}

\usepackage[dvipsnames]{xcolor}
\usepackage{soul}
\usepackage[colorlinks=true,urlcolor=MidnightBlue,citecolor=MidnightBlue,linkcolor=MidnightBlue]{hyperref}

\newcommand{\mathd}{\mathrm{d}}

\newcommand{\mathi}{\mathrm{i}}
\newcommand{\mathe}{\mathrm{e}}

\newcommand{\calP}{\mathcal{P}}

\newcommand{\oppsi}{\hat{\psi}}

\newcommand{\CFT}{\mathrm{CFT}}

\newcommand{\barJ}{\bar{J}}
\newcommand{\thesetof}[1]{\{#1\}}

\newcommand{\<}{\langle}
\renewcommand{\>}{\rangle}

\newcommand{\tr}{\mathrm{Tr}}

\newcommand\redsout{\bgroup\markoverwith{\textcolor{red}{\rule[0.5ex]{2pt}{0.4pt}}}\ULon}

\begin{document}

\preprint{APS/123-QED}

\title{Kac-Moody symmetries in one-dimensional bosonic systems}

\author{Wei Tang}
\author{Jutho Haegeman}
\affiliation{Department of Physics and Astronomy, Ghent University, Krijgslaan 281, 9000 Gent, Belgium}

\date{\today}

\begin{abstract}
In conformal field theories, when the conformal symmetry is enhanced by a global Lie group symmetry, the original Virasoro algebra can be extended to Kac-Moody algebra.  
In this paper, we extend the lattice construction of the Kac-Moody generators introduced in Wang \emph{et al.}, [\href{https://link.aps.org/doi/10.1103/PhysRevB.106.115111}{Phys.~Rev.~B. 106, 115111 (2022)}] to continuous systems and apply it to one-dimensional continuous boson systems.
We justify this microscopic construction of Kac-Moody generators in two ways.
First, through phenomenological bosonization, we express the microscopic construction in terms of the boson operators in the bosonization context, which can be related to the Kac-Moody generators in conformal field theories. 
Second, we study the behavior of the Kac-Moody generators in the integrable Lieb-Liniger model, and reveal its underlying particle-hole excitation picture through Bethe ansatz solutions.
Finally, we test the computation of the Kac-Moody generator in the continuous matrix product state simulations, paving the way for more challenging non-integrable systems.
\end{abstract}

\maketitle


\section{Introduction} 

Universality is a fundamental concept in the physical description of critical systems. 
For systems that are approaching a critical point or are already in a critical state, they exhibit similar behavior on large length scales, despite having different microscopic details.
In (1+1) dimensions, conformal field theory (CFT) serves as a powerful theoretical tool to compute the universal behavior of critical and near-critical systems~\cite{henkel-conformal-book-1999,francesco-conformal-book-2012}.

Given a critical system described by a CFT, one paramount task is to obtain the conformal data that fully determine the properties of the system at low energies and large length scales.
For example, it was discovered by Cardy and others~\cite{cardy-conformal-1984,blote-conformal-1986,cardy-logarithmic-1986,cardy-operator-1986,affleck-universal-1986} that the conformal data can be extracted from the low-energy states of a lattice system.
For lattice systems with periodic boundary conditions, the low-energy states of the lattice system can be regarded as approximations of the CFT states.
At the operator level, Koo and Saleur~\cite{koo-representations-1994} demonstrated that, in some integrable models, a lattice representation for the generators of the Virasoro algebra of the CFT can be established, known as the Koo-Saleur formula. 
This representation has recently also been successfully applied to non-integrable systems~\cite{milsted-extraction-2017}. 
Combined with periodic matrix product state techniques~\cite{pirvu-exploiting-2011,haegeman-variational-2012,pirvu-matrix-2012-b}, this formula leads to a series of systematic methods for extracting conformal data from lattice systems~\cite{milsted-extraction-2017,zou-conformal-2018,zou-conformal-2020,zou-emergence-2020}. 

In some CFTs, the conformal symmetries of the theory are enhanced by the presence of a larger symmetry. 
An important example is when the conformal symmetry is enhanced by a global Lie group symmetry.
In this case, the scaling operators are organized by an extension of the Virasoro algebra---the Kac-Moody algebra, which allows for a more compact characterization of the CFT~\cite{francesco-conformal-book-2012,blumenhagen-introduction-2009,mussardo-statistical-2020}. 
Recently, Ref.~\cite{wang-emergence-2022} proposed an approach to construct the generators of the Kac-Moody algebra as lattice operators in quantum spin chain systems.
Using a similar strategy as the Koo-Saleur formula for Virasoro generators, the lattice Kac-Moody generators for $\mathrm{U}(1)$ and $\mathrm{SU}(2)$ Kac-Moody algebras are constructed, which exhibit desired properties when acting on the low-energy states of the spin systems.
Furthermore, Ref.~\cite{yang-detecting-2022} presented a numerical approach to identify emergent symmetries at quantum critical points, including Kac-Moody symmetries, enabling the construction of the lattice realization of Kac-Moody generators to high accuracy.

In this paper, we extend the lattice construction of the Kac-Moody generator to (non-relativistic) continuous systems, which is referred to as the microscopic construction, using one-dimensional bosonic systems with particle-number conservation as a specific example. 
We justify this microscopic construction from two aspects. 
First, we represent it in terms of boson operators within the framework of phenomenological bosonization~\cite{haldane-effective-1981,cazalilla-bosonizing-2004}, which can be connected to the Kac-Moody generators in CFT.  
Second, we study the Kac-Moody generators in the specific example of the Lieb-Liniger model~\cite{lieb-exact-1963-a,lieb-exact-1963-b}, which is integrable.
Through the Bethe ansatz method~\cite{lieb-exact-1963-a,lieb-exact-1963-b,korepin-quantum-book-1997,gaudin-bethe-book-2014,yang-thermodynamics-1969,jiang-understanding-2015}, we compute the form factors of the microscopic construction of the Kac-Moody generators in the low-energy sector.
Our results demonstrate that the effect of the Kac-Moody generators on low-energy states can be regarded as particle-hole excitations in a certain fermionic picture, and the distribution of the fermion modes in momentum space is identical to the quantum number distribution of the Bethe ansatz wavefunctions.  
Finally, we also demonstrate the computation of Kac-Moody generators in the context of numerical simulations, where we employ the continuous matrix product state (cMPS) approach~\cite{verstraete-continuous-2010,haegeman-calculus-2013}.
We show that the cMPS simulations can correctly reproduce the behavior of form factors of the Kac-Moody generators, even without imposing the $\mathrm{U}(1)$ symmetry explicitly on the cMPS.
This provides a promising avenue for applying the microscopic construction of the Kac-Moody generators to more challenging problems where exact solutions are not available.

The remainder of the paper is organized as follows. 
In Sec.~\ref{sec:kacmoody-review}, following Ref.~\cite{wang-emergence-2022}, we briefly review the realization of the $\mathrm{U}(1)$ Kac-Moody generators on the lattice. 
In Sec.~\ref{sec:KM-1d-boson}, we discuss the microscopic construction of the Kac-Moody generators in the continuous one-dimensional models of interacting bosons and justify it with the phenomenological bosonization technique.  
In Sec.~\ref{sec:liebliniger}, we exemplify the Kac-Moody generator construction by studying the Lieb-Liniger model through the Bethe ansatz solution.
In Sec.~\ref{sec:cmps}, we discuss the numerical computation of the Kac-Moody generators in the context of cMPS simulations.
Section \ref{sec:conclusion-and-outlook} contains concluding remarks and outlooks. 

\section{$\mathbf{U(1)}$ Kac-Moody generators and its realization on the lattice} \label{sec:kacmoody-review}

In this section, following Ref.~\cite{wang-emergence-2022}, we briefly review the $\mathrm{U}(1)$ Kac-Moody algebra and its realization on the lattice.  

For a critical system described by a CFT with a global $\mathrm{U}(1)$ symmetry, its low-energy spectrum can be classified using the so-called $\mathrm{U}(1)$ Kac-Moody algebra. 
For a CFT with a global $\mathrm{U}(1)$ symmetry, we can define the $\mathrm{U}(1)$ charge $Q^{\CFT}=\int \mathd x \, q^{\CFT}(x)$.
The $\mathrm{U}(1)$ local current $q^{\CFT}(x)$ can be separated into the holomorphic part $J^{\CFT}(x)$ and anti-holomorphic part $\barJ^{\CFT}(x)$, i.e. $q^{\CFT}(x) = J^{\CFT}(x) + \barJ^{\CFT}(x)$.
Since both $J^{\CFT}(x)$ and $\barJ^{\CFT}(x)$ satisfy conservation laws, one can introduce another $\mathrm{U}(1)$ current $m^{\CFT}(x) = v(J^{\CFT}(x) - \barJ^{\CFT}(x))$ and another $\mathrm{U}(1)$ charge $M^{\CFT} = \int \mathd x \, m^{\CFT}(x)$, where $v$ is the velocity of the CFT.
Therefore, the global symmetry of the CFT is actually $\mathrm{U}(1)\times \mathrm{U}(1)$.
The two $\mathrm{U}(1)$ currents can be connected with each other by the conservation law 
\begin{equation}
    -\partial_x m^{\CFT}(x) = \mathi \partial_\tau q^{\CFT}(x) = \mathi [H^{\CFT}, q^{\CFT}(x)].
    \label{eq:u1charge_m}
\end{equation}

The Fourier modes $J_{m}$, $\barJ_{n}$ ($m, n \in \mathbb{Z}$) of $J^{\CFT}(x)$ and $\barJ^{\CFT}(x)$ satisfy the $\mathrm{U}(1)$ Kac-Moody algebra
\begin{align}
    [J_m^{\CFT}, J_n^{\CFT}] &= m \delta_{m+n, 0}, \\
    [\barJ_m^{\CFT}, \barJ_n^{\CFT}] &= m \delta_{m+n, 0}, \\
    [J_m^{\CFT}, \barJ_n^{\CFT}] &= 0.
\end{align}
The Virasoro generators can be expressed in terms of the Kac-Moody generators~\cite{blumenhagen-introduction-2009,mussardo-statistical-2020}
\begin{align}
    L^\CFT_m &= \frac{1}{2} :\sum_{n=-\infty}^{\infty} J^\CFT_{n+m} J^\CFT_{-n}: ,\\
    \bar{L}^\CFT_m &= \frac{1}{2} :\sum_{n=-\infty}^{\infty} \barJ^\CFT_{n+m} \barJ^\CFT_{-n}: ,
\end{align}
where $:\mathcal{O}: \equiv \mathcal{O} - \<\mathcal{O}\>_\mathrm{gs}$ represents the normal ordering of the operators.
We can then express the CFT Hamiltonian as
\begin{align}
    H^\CFT &= \frac{2\pi v}{L} \sum_{n=1}^\infty \left(J_{-n}^\CFT J_{n}^\CFT + \barJ_{-n}^\CFT \barJ_{n}^\CFT \right) \nonumber\\
           &\phantom{=} + \frac{2\pi v}{L}\left[\frac{1}{2}(J^\CFT_0 J^\CFT_0 + \barJ^\CFT_0 \barJ^\CFT_0) -\frac{c}{12} \right].
           \label{eq:CFT-Hamiltonian}
\end{align}
Here, $v$ is the velocity, $c=1$ is the central charge of the CFT, and we have used $H^\CFT = (2\pi v/L) (L^\CFT_0 + \bar{L}^\CFT_0 - c/12)$. 
A Kac-Moody primary state $|\alpha \>$ is defined by the following condition  
\begin{equation}
   J_m^{\CFT} |\alpha \> = 0, \barJ_m^{\CFT} |\alpha \> = 0 \; ( \forall m > 0).
\end{equation}
From a primary state $|\alpha\>$, one can construct descendant states by acting with $J_{-m}$'s and $\barJ_{-m}$'s ($m > 0$) on top of $|\alpha\>$ 
\begin{equation}
    J_{-1}^{k_1} J_{-2}^{k_2} \dots \barJ_{-1}^{\bar{k}_1} \barJ_{-2}^{\bar{k}_2} \dots |\alpha\>,
\end{equation}
where $k_1, \bar{k}_1, k_2, \bar{k}_2 \dots \geq 0$.
These descendant states, together with the primary state $|\alpha\>$, constitute the so-called Kac-Moody tower.

On the lattice, for conformal critical systems with a global $\mathrm{U}(1)$ symmetry, one can identify the $\mathrm{U}(1)$ charge $Q=\sum_j q_j$ with the $\mathrm{U}(1)$ charge $Q^{\CFT}$ in the CFT, and then $q_j$ corresponds to $q^{\CFT}(x)$.
Similarly, according to Eq.~\eqref{eq:u1charge_m}, one can introduce a quantity $m_j$ defined on the lattice satisfying 
\begin{equation}
    m_{j+1} - m_{j} = \mathi [H, q_j].
    \label{eq:mj_lat}
\end{equation}
We associate $m_{j}$ to $m^{\CFT}(x)$. 
Along this line, we introduce the lattice realizations $J_n$, $\barJ_n$ of the Kac-Moody generators by performing the Fourier transformation of $q_j$ and $m_j$, i.e.,
\begin{align}
    J_n &= \sum_j^N \mathe^{-2 \pi \mathi j n / N} \frac{q_j + m_j / v}{2}, \label{eq:Jn_lat} \\ 
    \barJ_n &= \sum_j^N \mathe^{2 \pi \mathi j n / N} \frac{q_j - m_j / v}{2}, \label{eq:barJn_lat} 
\end{align}
where the velocity $v$ corresponds to the velocity of low-energy excitations in the lattice system.

Compared to $J^{\CFT}_n$ and $\barJ^{\CFT}_n$, the operators $J_n$ and $\barJ_n$ constructed in lattice models contain contributions of irrelevant terms at short length scales, and thus do not satisfy the Kac-Moody algebra. 
The expectation is that, when applied on the low-energy states, the contributions from those irrelevant terms are negligible, and $J_n$ and $\barJ_n$ will have the same matrix elements in low-energy subspace as the Kac-Moody generators $J^{\CFT}_n$ and $\barJ^{\CFT}_n$ in the CFT.

\section{Kac-Moody generator in the one-dimensional bosonic systems} \label{sec:KM-1d-boson}

In this section, we discuss the realization of Kac-Moody generators in one-dimensional (non-relativistic) continuous bosonic systems with particle-number conservation, which can be described by the following Hamiltonian
\begin{equation}
    H = \int_0^L \mathd x \partial_x \psi^\dagger (x) \partial_x \psi(x) + \int_0^L \mathd x \int_0^L \mathd x' u(x-x') \rho(x) \rho(x'),
    \label{eq:general-hamiltonian}
\end{equation}
where $\rho(x) = \psi^\dagger(x) \psi(x)$ is the density operator, $u(x-x') = u(|x-x'|)$ is the interaction potential function, and $L$ is the system size. 
In condensed matter theories, the Kac-Moody symmetry in this system is incorporated in the theoretical framework known as Luttinger liquid theory, and the Kac-Moody algebra is reformulated in the bosonic nature of the low energy excitations~\cite{haldane-effective-1981,haldane-luttinger-1981,ludwig-methods-1995,vondelft-bosonization-1998}.
Hence, in our discussion, we will first give the microscopic construction of the Kac-Moody generator in the bosonic system \eqref{eq:general-hamiltonian}, and then demonstrate its correctness using the phenomenological bosonization technique~\cite{haldane-effective-1981,cazalilla-bosonizing-2004}. 

To obtain the microscopic construction of the Kac-Moody generators, it is straightforward to generalize the lattice construction \eqref{eq:mj_lat}, \eqref{eq:Jn_lat}, \eqref{eq:barJn_lat} to continuous systems. 
The $\mathrm{U}(1)$ symmetry of the system corresponds to the particle-number conservation, and the $\mathrm{U}(1)$ charge is thus the total particle number $N = \int \mathd x\,\rho(x)$.
Suppose the particle density of the ground state is $\rho_0 = N_0 / L$, we identify the local particle density fluctuation $\Delta\rho(x) = \rho(x) - \rho_0$ with $q^{\CFT}(x)$ in the CFT, and note that 
\begin{equation}
    \mathi [H, \Delta{\rho}(x)] = \mathi \left[ \psi^\dagger(x) \partial_x^2 \psi(x) - \partial_x^2 \psi^\dagger(x) \psi(x) \right]
                     = -\partial_x j(x),
\end{equation}
where we have introduced the density current operator
\begin{equation}
    j(x) = -\mathi \left[ \psi^\dagger(x) \partial_x \psi(x) - \partial_x \psi^\dagger(x) \psi(x) \right]. 
    \label{eq:density-current-operator}
\end{equation}
We then identify $j(x)$ with $m^{\CFT}(x)$ in the CFT.
Therefore, according to the discussion in Sec.~\ref{sec:kacmoody-review}, in a system of length $L$ with periodic boundary conditions, we can construct the microscopic realizations of Kac-Moody generators $J_n$ and $\barJ_n$ as 
\begin{align}
    J_n &= \int_0^L \mathd x \, \mathe^{-2\pi n \mathi x /L} \frac{\Delta\rho(x) + j(x) / v}{2}, \label{eq:J-liebliniger}\\
    \barJ_n &= \int_0^L \mathd x \, \mathe^{2\pi n \mathi x /L} \frac{\Delta\rho(x) - j(x) / v}{2}.
    \label{eq:barJ-liebliniger}     
\end{align}

Next, following Refs.~\cite{haldane-effective-1981,cazalilla-bosonizing-2004}, we introduce the underlying Luttinger liquid description of the model \eqref{eq:general-hamiltonian} using phenomenological bosonization. 
To start with, we employ the density-phase representation of the boson operator  
\begin{equation}
    \psi^\dagger (x) = \sqrt{\rho(x)} \mathe^{- \mathi \phi (x)},\; \psi(x) = \mathe^{\mathi \phi(x)} \sqrt{\rho(x)},
    \label{eq:density-phase-repr}
\end{equation}
where $\rho(x)$ and $\phi(x)$ are Hermitian operators which represent the boson density and phase, respectively.
We introduce an auxiliary field $\Theta (x)$ and represent the density operator as 
\begin{equation}
    \rho(x) = [\partial_x \Theta (x)] \sum_{n=-\infty}^{\infty} \delta(\Theta(x) - n\pi). \label{eq:rho-theta-first-quantized}
\end{equation}
Equation \eqref{eq:rho-theta-first-quantized} is equivalent to the first quantized form of the density operator $\rho(x) = \sum_n \delta(x - x_n)$ provided that the particle positions $\{x_n\}$ satisfy $\Theta (x_n) = n\pi$, where we have used the relation $\delta [f(x)] = \delta(x - x_0) / |f'(x_0)|$, $f(x_0) = 0$. 

To study the low-energy physics of the system, we henceforth take both $\Theta(x)$ and $\phi(x)$ as slowly varying fields by coarse-graining them over a length scale $l \gg \rho_0^{-1}$.
Moreover, using the Poisson's summation formula, we rewrite Eq.~\eqref{eq:rho-theta-first-quantized} in a more useful form 
\begin{equation}
    \rho(x) = \frac{1}{\pi} [\partial_x \Theta (x)] \sum_{m=-\infty}^{\infty} \mathe^{2m \mathi \Theta(x)}. \label{eq:rho-theta-poisson-sum}
\end{equation}
Equation \eqref{eq:rho-theta-poisson-sum} separates the density fluctuations of different length scales. 
The $m=0$ term describe the density fluctuations of length scale $l \gg \rho_0^{-1}$, whereas the terms with $m \neq 0$ describe density fluctuations of length scale $(m\rho_0)^{-1}$.
Therefore, at the long-wave-length limit, it suffices to only keep the $m=0$ term in Eq.~\eqref{eq:rho-theta-poisson-sum}, which yields
\begin{equation}
    \rho(x) \approx \frac{1}{\pi} \partial_x \Theta(x).
    \label{eq:rho-long-wave-length}
\end{equation} 
Combining Eq.~\eqref{eq:rho-long-wave-length} and Eqs.~\eqref{eq:density-current-operator} and \eqref{eq:density-phase-repr}, we obtain an approximation for the density current operator  
\begin{equation}
    j(x) \approx 2 \rho_0 \partial_x \phi(x). 
    \label{eq:j-long-wave-length}
\end{equation}

By substituting Eqs.~\eqref{eq:density-phase-repr} and \eqref{eq:rho-long-wave-length} into the Hamiltonian \eqref{eq:general-hamiltonian} and keeping only the leading terms, one can obtain the long-wave-length effective Hamiltonian 
\begin{equation}
    H_{\mathrm{eff}} = \frac{1}{2\pi} \int_0^L \mathd x \left[ v_J (\partial_x \phi(x))^2 + v_N (\partial_x \Theta(x) - \pi \rho_0)^2 \right],
    \label{eq:effective-hamiltonian-1}
\end{equation}
where $v_J = 2\pi \rho_0$ is the phase stiffness, and $v_N$ is the density stiffness.  
It is customary in the bosonization literature to introduce the field $\theta(x) = \Theta(x) - \pi \rho_0 x$, the Luttinger parameter $K=\sqrt{v_J / v_N}$, and the velocity $v = \sqrt{v_N v_J}$. We then rewrite Eq.~\eqref{eq:effective-hamiltonian-1} as 
\begin{equation}
    H_{\mathrm{eff}} = \frac{v}{2 \pi} \int_0^L \mathd x \left[ K (\partial_x \phi(x))^2 + \frac{1}{K} (\partial_x \theta(x))^2 \right].
    \label{eq:effective-hamiltonian-K}
\end{equation}
Here, the velocity $v$ describes the velocity of the low-energy excitations, and thus is identical to the velocity that appears in Eqs.~\eqref{eq:J-liebliniger} and \eqref{eq:barJ-liebliniger}. 
The Luttinger parameter $K$ is related to the strength of the quantum fluctuations. 
These two parameters fully characterize the Luttinger liquid theory.

To properly diagonalize the effective Hamiltonian \eqref{eq:effective-hamiltonian-K}, we employ the following mode expansion~\cite{haldane-effective-1981}
\begin{align}
\Theta(x) &= \theta_0 + \frac{\pi N x}{L} - \mathi \sum_{q\neq 0} \left| \frac{\pi K}{2 q L} \right|^{\frac{1}{2}} \mathrm{sgn}(q) \mathe^{\mathi q x} (b_q^\dagger + b_{-q}), \label{eq:Theta-expansion}\\
\phi(x) &= \phi_0 + \frac{\pi J x}{L} -\mathi \sum_{q\neq 0} \left| \frac{\pi}{2 q L K} \right|^{\frac{1}{2}} \mathe^{\mathi q x} (b^\dagger_q - b_{-q}), \label{eq:phi-expansion}
\end{align}
where we have assumed that the system obeys the periodic boundary condition, and $q = \pm 2 \pi n / L, n\in \mathbb{N}^+$. 
The operators $b_q$ and $b^\dagger_q$ are boson operators satisfying $[b_q, b_{q'}^\dagger]=\delta_{q, q'}$ which describe the low-energy collective excitations.
The operators $N$ and $J$ corresponds to the total particle number and total current respectively, which, together with the zero modes $\theta_0, \phi_0$, satisfy the following commutation relations: $[N, \mathe^{-\mathi \phi_0}] = \mathe^{-\mathi \phi_0}$, $[J, \mathe^{-\mathi \theta_0}] = \mathe^{-\mathi \theta_0}$, and $[N, J] = [\theta_0, \phi_0] = 0$. 
These commutation relations will lead to the correct commutation relation between the field operators $[\partial_x \Theta(x), \phi(x')] = \mathi \pi \delta(x - x')$, which follows from the commutation relations of the boson fields $\psi(x)$ and $\psi^\dagger(x)$. 

By substituting mode expansions \eqref{eq:Theta-expansion} and \eqref{eq:phi-expansion} into the effective Hamiltonian \eqref{eq:effective-hamiltonian-K}, we get  
\begin{equation}
    H_{\mathrm{eff}} = \sum_{q \neq 0} v |q| b_q^\dagger b_q + \frac{\pi v}{2LK}(N-N_0)^2 + \frac{\pi v K}{2L} J^2 + \mathrm{const}. 
    \label{eq:diagonalized-effective-H}
\end{equation}
One can easily verify that the spectrum given by Eq.~\eqref{eq:diagonalized-effective-H} is identical to the spectrum of the CFT Hamiltonian \eqref{eq:CFT-Hamiltonian}, provided that we make the following identifications
\begin{align}
    J^{\CFT}_n \sim \sqrt{n} b_{2\pi n / L}&,\; J^{\CFT}_{-n} \sim \sqrt{n} b^\dagger_{2\pi n / L}, \label{eq:identification-R} \\
    \barJ^{\CFT}_n \sim \sqrt{n} b_{-2\pi n / L}&,\; \barJ^{\CFT}_{-n} \sim \sqrt{n} b^\dagger_{-2\pi n / L}, \label{eq:identification-L} \\
    J_0^\CFT + \barJ_0^\CFT \sim \frac{N-N_0}{\sqrt{K}}&,\; J_0^\CFT - \barJ_0^\CFT \sim \sqrt{K} J,
\end{align}
where $n$ is a positive integer.

Regarding the Kac-Moody generators, as we only apply the Kac-Moody generators to states with low energies and small momenta, it suffices to employ the long-wave-length approximation.
Combining Eqs.~\eqref{eq:rho-long-wave-length}, \eqref{eq:j-long-wave-length}, \eqref{eq:Theta-expansion} and \eqref{eq:phi-expansion}, we get 
\begin{align}
    \Delta\rho(x) &\approx \frac{N-N_0}{L} + \frac{\sqrt{K}}{L} \sum_{q\neq 0} \sqrt{\frac{|q| L}{2\pi}} \mathe^{\mathi q x}(b_q^\dagger + b_{-q}), \label{eq:rho-expansion} \\
    j(x) &\approx \frac{v K J}{L} + \frac{v \sqrt{K}}{L} \sum_{q\neq 0} \sqrt{\frac{|q| L}{2\pi}} \mathrm{sgn}(q) \mathe^{\mathi q x} (b_q^\dagger - b_{-q}), \label{eq:j-expansion}
\end{align}
where in Eq.~\eqref{eq:j-expansion} we have used $2\pi \rho_0 = v_J = v K$. 
By substituting the mode expansions \eqref{eq:rho-expansion} and \eqref{eq:j-expansion} into Eqs.~\eqref{eq:J-liebliniger}, \eqref{eq:barJ-liebliniger}, and combining Eqs.~\eqref{eq:identification-R} and \eqref{eq:identification-L}, it becomes clear that the microscopic constructions \eqref{eq:J-liebliniger}, \eqref{eq:barJ-liebliniger} are identical to the sought Kac-Moody generators in the CFT up to an overall factor $\sqrt{K}$, i.e., 
\begin{equation}
    J_n^\CFT = J_n / \sqrt{K}, \barJ_n^\CFT = \barJ_n / \sqrt{K} , \label{eq:JCFT_and_J}
\end{equation}
for any integer $n$. 
In the following contexts, we will absorb the factor $1/\sqrt{K}$ into $J_n$ and $\barJ_{n}$ for the convenience of discussion. 

\section{Application to the Lieb-Liniger model} \label{sec:liebliniger}
As a specific example, we examine the microscopic construction of the Kac-Moody generators in the bosons interacting with a zero-range potential, i.e., the Lieb-Liniger model~\cite{lieb-exact-1963-a,lieb-exact-1963-b} in this section. 
The Lieb-Liniger Hamiltonian reads 
\begin{equation}
    \hat{H} = \int_0^L \mathd x \left[\partial_x\psi^\dagger(x) \partial_x\psi(x) + c \psi^\dagger(x) \psi^\dagger(x) \psi(x) \psi(x) \right],
    \label{eq:lieb-liniger}
\end{equation}
where $\mu$ is the chemical potential and $c>0$ is the interaction strength. 

As an integrable system, the Lieb-Liniger model is one of the prototypical models which can be exactly solved by Bethe ansatz and plays a central role in the early development of Bethe ansatz method~\cite{korepin-quantum-book-1997,gaudin-bethe-book-2014,lieb-exact-1963-a,lieb-exact-1963-b,yang-thermodynamics-1969,jiang-understanding-2015}.
In the following, using the Bethe ansatz approach, we will obtain the eigenstates of the Lieb-Liniger Hamiltonian \eqref{eq:lieb-liniger} represented as Bethe states, and then study the behavior of $J_n$ and $\barJ_n$ when acting with them on the low-energy eigenstates.  

\paragraph{Bethe wavefunctions} The Bethe wavefunction with $N$ bosons is expressed as 
\begin{equation} \label{eq:betheWaveFunc}
     \psi_{\thesetof{\lambda_j}}(\boldsymbol{x}) = \< \boldsymbol{x} | \thesetof{\lambda_j} \>= \sum_{\calP} a(\calP) \exp(\mathi \sum_{j=1}^N \lambda_{\calP(j)} x_j),
\end{equation}
where the parameters $\thesetof{\lambda_j} = \thesetof{\lambda_1, \lambda_2, \ldots, \lambda_N}$ are called the quasimomenta, and $\calP$ represents a permutation of the quasimomenta. 
For the Lieb-Liniger model, the coefficient $a(\calP)$ is given by 
\begin{equation}
    a(\calP) = \prod_{j<k, \calP(j) > \calP(k)}\frac{\mathi (\lambda_j - \lambda_k) - c}{\mathi (\lambda_j - \lambda_k) + c}.
\end{equation}
Next, we impose the periodic boundary condition, yielding the following Bethe equations 
\begin{equation} \label{eq:betheEqs}
    \exp (\mathi \lambda_j L) = (-1)^{N-1} \prod_{k=1}^N \frac{c - \mathi (\lambda_j - \lambda_k)}{c + \mathi (\lambda_j - \lambda_k)}, j=1, \ldots, N.
\end{equation}
For the Lieb-Liniger model, one can show that the possible quasimomenta can only take real values~\cite{korepin-quantum-book-1997}. 
The quasimomenta $\{\lambda_j\}$ can be further associated with $N$ quantum numbers $\{I_j\}$ which satisfy~\cite{yang-thermodynamics-1969} 
\begin{equation}
    \lambda_j = \frac{2 \pi I_j + \sum_{k=1}^N \theta_{j k}}{L},
    \label{eq:I-lambda-ba}
\end{equation} 
where $\theta_{j k} = -2 \arctan[(\lambda_j - \lambda_k)/c]$. 
The quantum numbers ${I_j}$ take integer values when $N$ is odd, and half-integer values when $N$ is even.
The distribution of the quantum numbers $\{I_j\}$ completely determines the quasimomenta $\lambda_j$ and thus the eigenstate $|\{\lambda_j\}\>$.
In practice, one has the freedom to choose a set of $\{I_j\}$, and the corresponding quasimomenta $\{\lambda_j\}$ can be solved by combining Eqs.~\eqref{eq:betheEqs} and \eqref{eq:I-lambda-ba}.
For a given set of quasimomenta $\thesetof{\lambda_j}$, the total momentum and energy of the state $\psi_{\thesetof{\lambda_j}} (\boldsymbol{x})$ are given by 
\begin{equation}
    P_{\{\lambda_j\}} = \sum_{j=1}^N \lambda_j = \sum_{j=1}^N \frac{2\pi I_j}{L}, \; 
    E_{\{\lambda_j\}} = \sum_{j=1}^N \lambda_j^2 .\label{eq:PandE}
\end{equation}

\begin{figure}[!htb]
    \centering
    \resizebox{0.9\columnwidth}{!}{\includegraphics{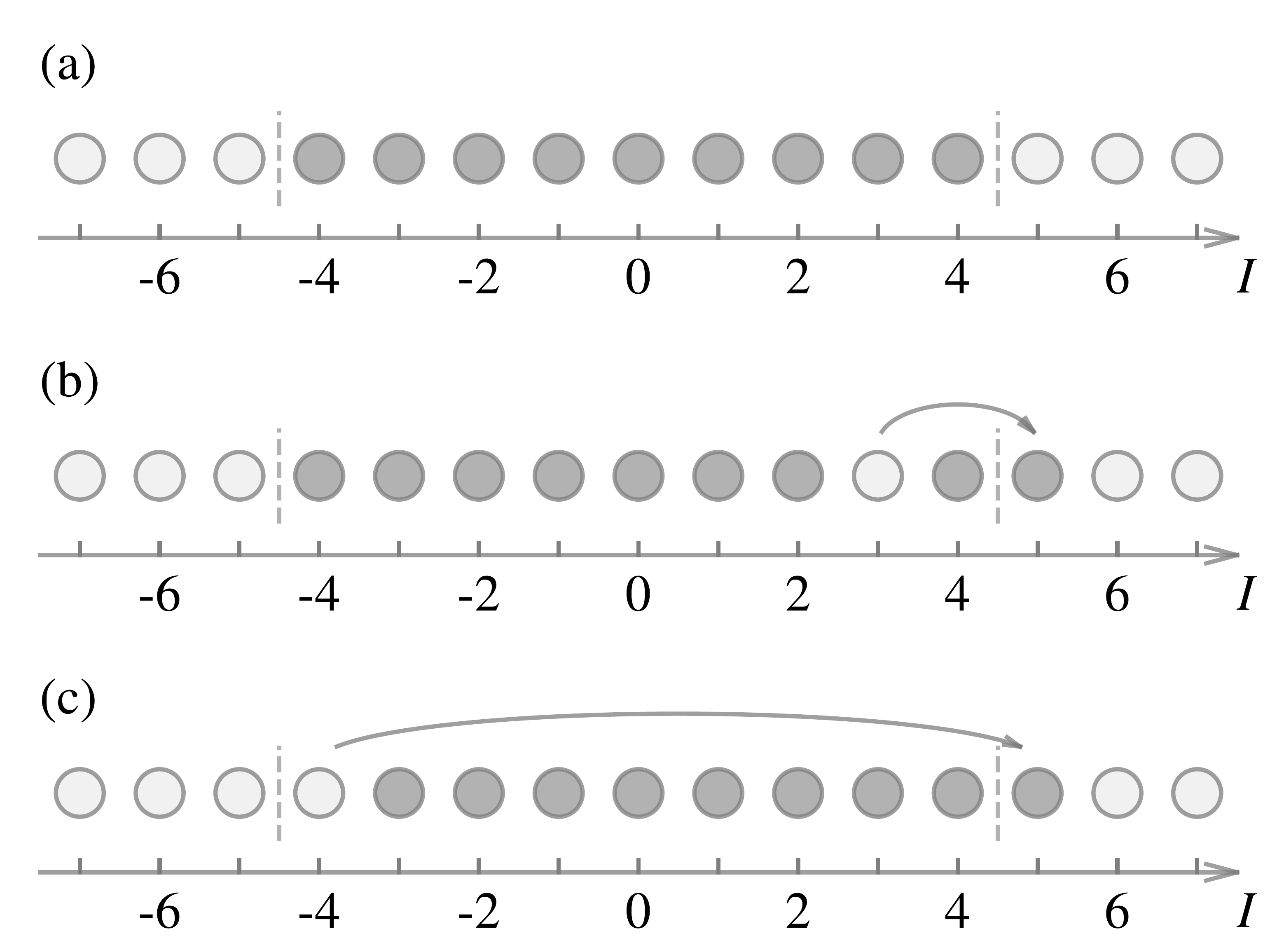}}
    \caption{Schematic diagrams showing the distributions of the quantum numbers $\thesetof{I_j}$ for (a) the ground state, (b) the low-energy state with particle-hole excitation, and (c) the scattering from one branch to another. }
    \label{fig:fermi-sea}
  \end{figure}

These quantum numbers cannot coincide with each other, otherwise the wavefunction vanishes~\cite{korepin-quantum-book-1997}.
For the ground state, the distribution of the quantum numbers $\{I_j\}$ resembles a Fermi sea, where the modes with the smallest absolute values are fully occupied [see Fig.~\ref{fig:fermi-sea} (a)]. 
The low-energy excited states with the same particle number can be generated from the ground state by introducing particle-hole excitations near the Fermi surface, or by moving one occupied mode from one branch to another [see Fig.~\ref{fig:fermi-sea} (b) and (c), respectively].
From Eq.~\eqref{eq:PandE}, it becomes clear that scattering between the different branches gives rise to low-energy excitations with length scales $(\rho_0)^{-1}$ or even smaller, which have been excluded by our previous analysis in Sec.~\ref{sec:KM-1d-boson}~[c.f. Eq.~\eqref{eq:rho-theta-poisson-sum}]. In contrast, the low-energy particle-hole excitations near the Fermi surface all have a length scale much larger than $(\rho_0)^{-1}$. 
Indeed, it is well known in the bosonization literature~\cite{ludwig-methods-1995,vondelft-bosonization-1998} that the low-energy bosonic excitations (and thus the Kac-Moody generators) can be interpreted as particle-hole excitations in a certain fermionic picture.
More specifically, for a positive integer $n \ll N_0$, we have
\begin{align}
    J_{-n} \approx \sum_{p=k_F-\Lambda}^{k_F+\Lambda} f^\dagger_{p+q} f_{p}&, \; 
    J_{+n} \approx \sum_{p=k_F-\Lambda}^{k_F+\Lambda} f^\dagger_{p-q} f_{p}, \label{eq:b-as-fs} \\
    \barJ_{-n} \approx \sum_{p=-k_F+\Lambda}^{-k_F-\Lambda} f^\dagger_{p-q} f_{p}&, \; 
    \barJ_{+n} \approx \sum_{p=-k_F+\Lambda}^{-k_F-\Lambda} f^\dagger_{p+q} f_{p}\label{eq:b-as-fs2},
\end{align}
where $f_p$ and $f_p^\dagger$ are fermion operators in the fermionic picture mentioned above, $q = 2 \pi n / L$, $k_F = 2\pi \rho_0/L$, and $\Lambda$ is the momentum cut-off.
Along this line, one can infer that the distribution of the fermion modes in the momentum space should be identical to the distribution of the quantum numbers $\thesetof{I_j}$, which will be confirmed through the calculations of the form factors in the following.

\paragraph{Form factors of Kac-Moody generators} To check the behavior of $J_n$ and $\barJ_n$ defined in Eqs.~\eqref{eq:J-liebliniger} and \eqref{eq:barJ-liebliniger}, we compute the matrix elements for the low-energy states, 
\begin{equation}
    C_{\mu, \lambda}[X] \equiv \frac{\< \thesetof{\mu_j} | X | \thesetof{\lambda_j} \>}{\sqrt{\<\thesetof{\mu_j}|\thesetof{\mu_j}\> \< \thesetof{\lambda_j}|\thesetof{\lambda_j}\>}} ,
    \label{eq:kac-moody-matrix-element}
\end{equation} 
where $X$ is an operator, and $|\thesetof{\mu_j}\>$ and $|\thesetof{\lambda_j}\>$ are low-energy eigenstates of the systems represented as Bethe states. 
To compute $C_{\mu, \lambda}[J_n]$ and $C_{\mu, \lambda}[\barJ_n]$, we plug Eqs.~\eqref{eq:J-liebliniger} and \eqref{eq:barJ-liebliniger} into Eq.~\eqref{eq:kac-moody-matrix-element}, and recall that we have absorbed an additional factor $1/\sqrt{K}$ into $J_n$ and $\barJ_n$ [c.f. Eq.~\eqref{eq:JCFT_and_J}].  
Since the momentum of a Bethe state $|\thesetof{\mu_j}\>$ is given by $p_{\thesetof{\mu_j}} = \sum_j^N \mu_j$ [see Eq.~\eqref{eq:PandE}], we can shift the operators $\rho(x)$ and $j(x)$ to position $x=0$ by inserting spatial translation operators in the expression. 
Along this line, we get 
\begin{align}
    C_{\mu, \lambda}[J_{-n}] &= \frac{L}{\sqrt{K}} \delta \left(p_{\thesetof{\mu_j}} - p_{\thesetof{\lambda_j}}, \frac{2\pi n}{L}\right) \left( C_{\mu, \lambda}[\rho(0)] + \frac{C_{\mu, \lambda}[j(0)]}{v} \right), \label{eq:J_form_factors_ba_computation}\\
    C_{\mu, \lambda}[\barJ_{-n}] &= \frac{L}{\sqrt{K}} \delta \left(p_{\thesetof{\lambda_j}} - p_{\thesetof{\mu_j}}, \frac{2\pi n}{L}\right) \left( C_{\mu, \lambda}[\rho(0)] - \frac{C_{\mu, \lambda}[j(0)]}{v} \right), \label{eq:barJ_form_factors_ba_computation}
\end{align}
where $\delta(p_1, p_2) = 1$ only when $p_1=p_2$, and equals to zero otherwise.
The velocity $v$ can be obtained from the low-energy spectrum, and the Luttinger parameter $K$ can be determined using the relation $v K = 2\pi \rho_0$.
The form factors $C_{\mu, \lambda}[\rho(0)]$ and $C_{\mu, \lambda}[j(0)]$ can be calculated with algebraic Bethe ansatz methods~\cite{slavnov-nonequal-1990,de-nardis-density-2015}. 
We list the expressions of these form factors in Appendix \ref{app:form-factors-ba}. 

In Fig.~\ref{fig:ba-spect}, we show the energy spectrum of a Lieb-Liniger model with $c=1$, $L=64$. 
The particle number is fixed to be $N=64$.
The eigenstates shown in the energy spectrum all correspond to different particle-hole scattering modes. 
In the spectrum, we can choose a low-energy state $|\psi_\mathrm{i}\>$ as an initial state.
By calculating the form factors, we can find out the states in the low energy spectrum which can be obtained by acting with the Kac-Moody generator $J_n$ on $|\psi_\mathrm{i}\>$.
Here, $n= \pm 1, \pm 2, \ldots$, and different choices of $n$ lead to states with different momenta.
In Fig.~\ref{fig:ba-spect}, we choose $|\psi_\mathrm{i} \>$ as the ground state $|\psi_0\>$ and the first excited state $J_{-1} |\psi_0\>$, respectively.  
Since the Kac-Moody generators consist of superpositions of particle-hole excitations [c.f. Eqs.~\eqref{eq:b-as-fs} and \eqref{eq:b-as-fs2}], at high energy levels, multiple eigenstates have non-zero overlap with $J_n |\psi_\mathrm{i}\>$. 
Moreover, the relative signs of these overlaps, which are also shown in Fig.~\ref{fig:ba-spect}, can be understood in terms of the fermionic nature of the particles in the particle-hole scattering picture.

\begin{figure}[!htb]
    \centering
    \resizebox{\columnwidth}{!}{\includegraphics{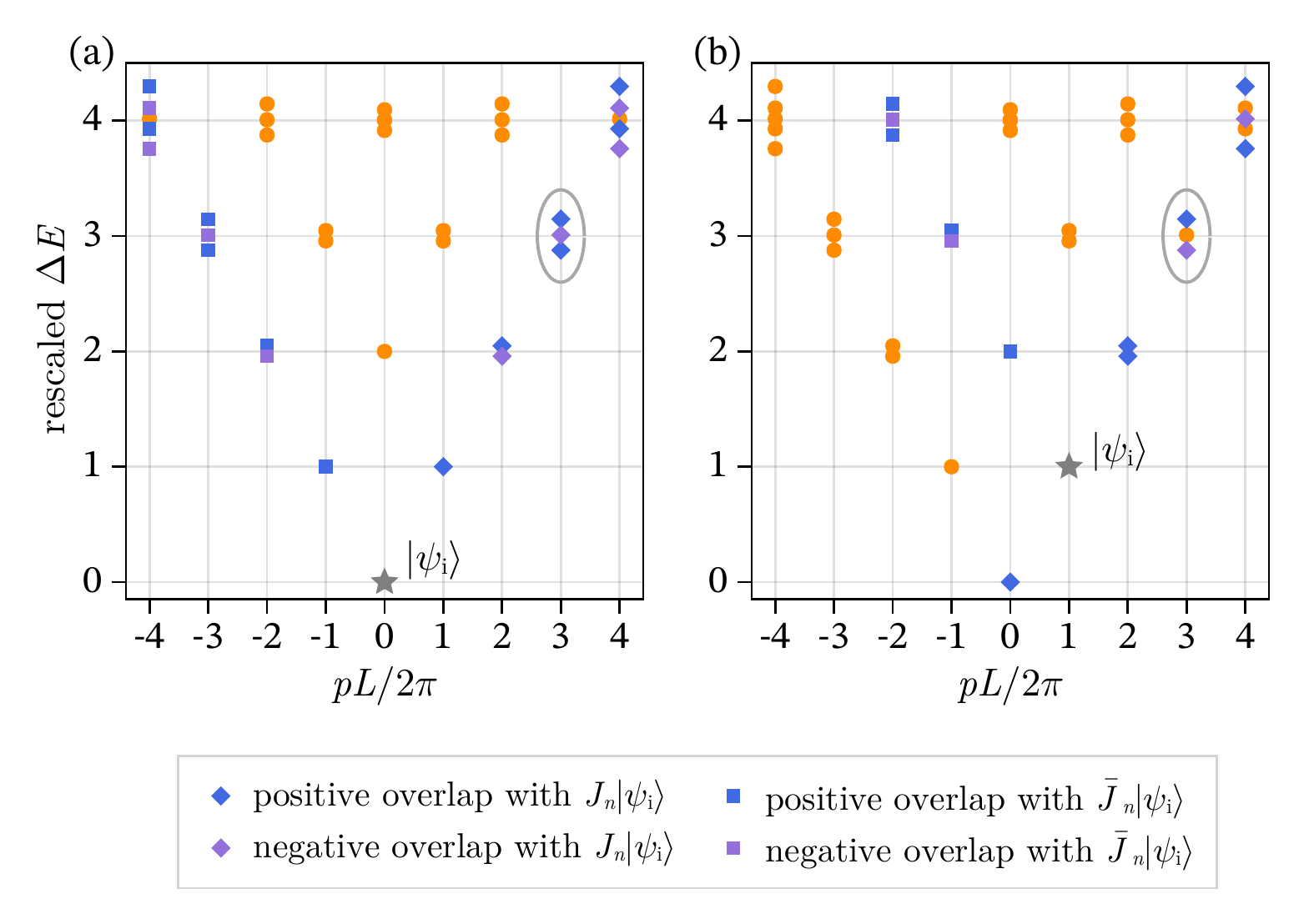}}
    \caption{Energy spectrum of the Lieb-Liniger model with $c=1$, $L=64$ and particle number $N=64$. The states can be mapped from $|\psi_i\>$ (denoted by the gray star) with $J_n$'s and $\bar{J}_n$'s are marked with different shapes and colors. The initial states are chosen as (a) the ground state $|\psi_0\>$ and (b) the first excited state $J_{-1}|\psi_0\>$. The states circled by the gray ellipse will be further discussed. }
    \label{fig:ba-spect}
  \end{figure}

\begin{figure}[!htb]
    \centering
    \resizebox{0.9\columnwidth}{!}{\includegraphics{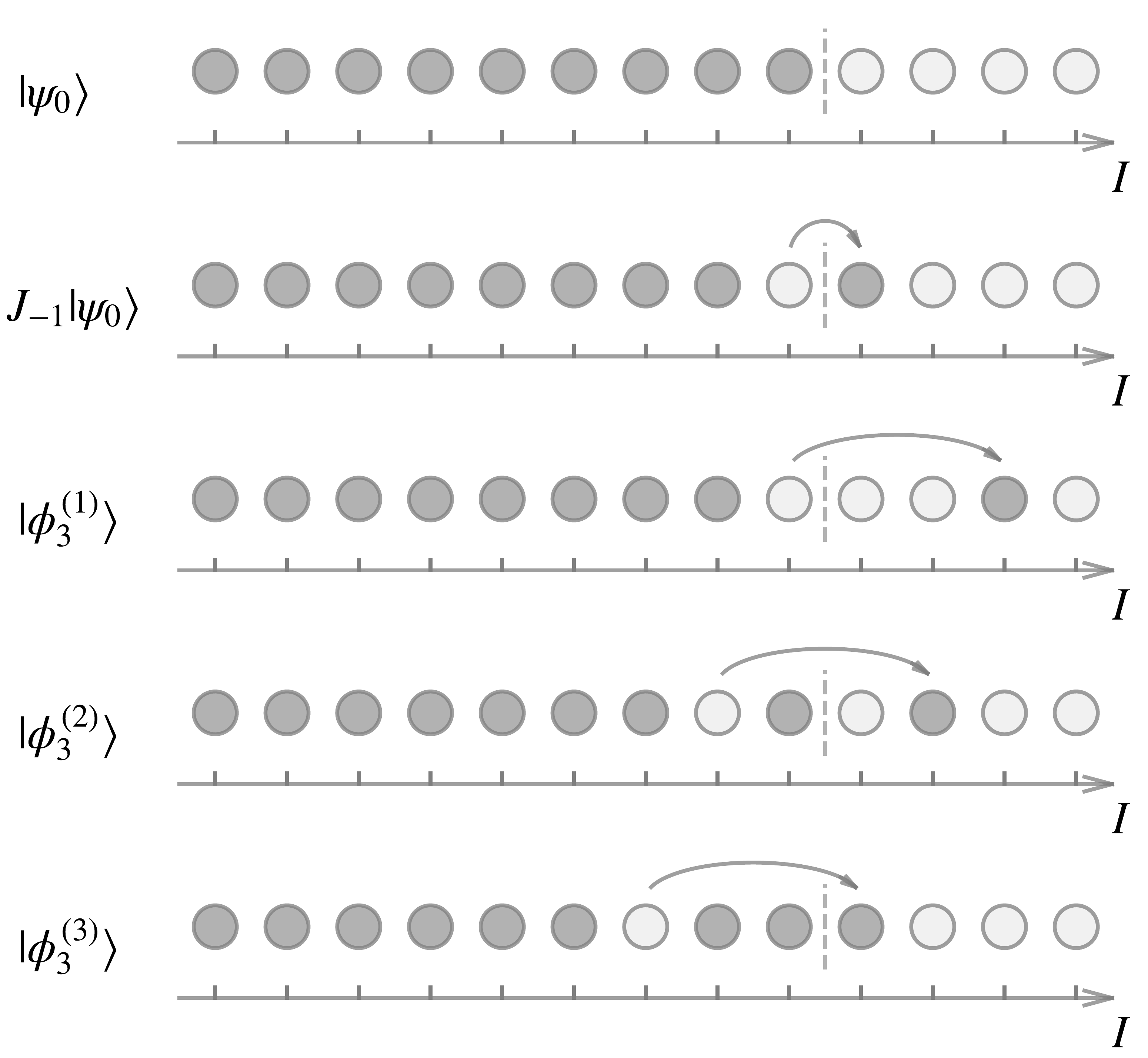}}
    \caption{The quantum number distribution of states $|\psi_0\>$, $J_{-1} |\psi_0\>$, $| \phi_3^{(1)} \>$, $| \phi_3^{(2)} \>$, and $| \phi_3^{(3)} \>$. The dashed line represent the Fermi surface.} 
    \label{fig:ba-ph3}
  \end{figure}

By looking into the quantum number configuration of the excited states, we can verify that the fermion modes in Eqs.~\eqref{eq:b-as-fs} and \eqref{eq:b-as-fs2} indeed correspond to the quantum numbers $\thesetof{I_j}$. 
As an example, we look into the states circled by the gray ellipse in Fig.~\ref{fig:ba-spect} for more details.
In Fig.~\ref{fig:ba-spect} (a), we have chosen the initial state as the ground state $|\psi_0\>$.
The three states $| \phi_3^{(1)} \>$, $| \phi_3^{(2)} \>$, and $| \phi_3^{(3)} \>$ circled by the gray ellipse can all be mapped from the ground state via the Kac-Moody generator $J_{-3}$. 
More concretely, we have 
\begin{equation}
    J_{-3} |\psi_0\> \approx | \phi_3^{(1)} \> - | \phi_3^{(2)} \> + | \phi_3^{(3)} \>.
\end{equation}
The quantum number distributions of states $| \phi_3^{(1)} \>$, $| \phi_3^{(2)} \>$, and $| \phi_3^{(3)} \>$ are shown in Fig.~\ref{fig:ba-ph3}. 
From Fig.~\ref{fig:ba-ph3}, it also becomes clear that the negative sign in front of $|\phi_3^{(2)}\>$ comes from the fermion commutation relation [c.f. Eqs.~\eqref{eq:b-as-fs} and \eqref{eq:b-as-fs2}].
In Fig.~\ref{fig:ba-spect} (b), the initial state is chosen as $J_{-1} |\psi_0\>$.
According to the momentum difference, the states in the ellipse can only be reached by acting with $J_{-2}$ on $J_{-1}|\psi_0\>$
\begin{equation}
    J_{-2} J_{-1} |\psi_0\> \approx | \phi_3^{(1)} \> - | \phi_3^{(3)} \>.
    \label{eq:J-2J-1psi0}    
\end{equation}
Here, $|\phi_3^{(2)}\>$ is excluded since it requires more than one particle-hole scattering processes to be obtained from $J_{-1}|\psi_0\>$. 
Moreover, a negative sign also appears in front of $| \phi_3^{(3)} \>$ in Eq.~\eqref{eq:J-2J-1psi0}, which, again, comes from the fermion commutation relation. 
As shown in Fig.~\ref{fig:ba-ph3}, the form factor computations results are indeed consistent with the discussions here.

\begin{figure}[!htb]
    \centering
    \resizebox{\columnwidth}{!}{\includegraphics{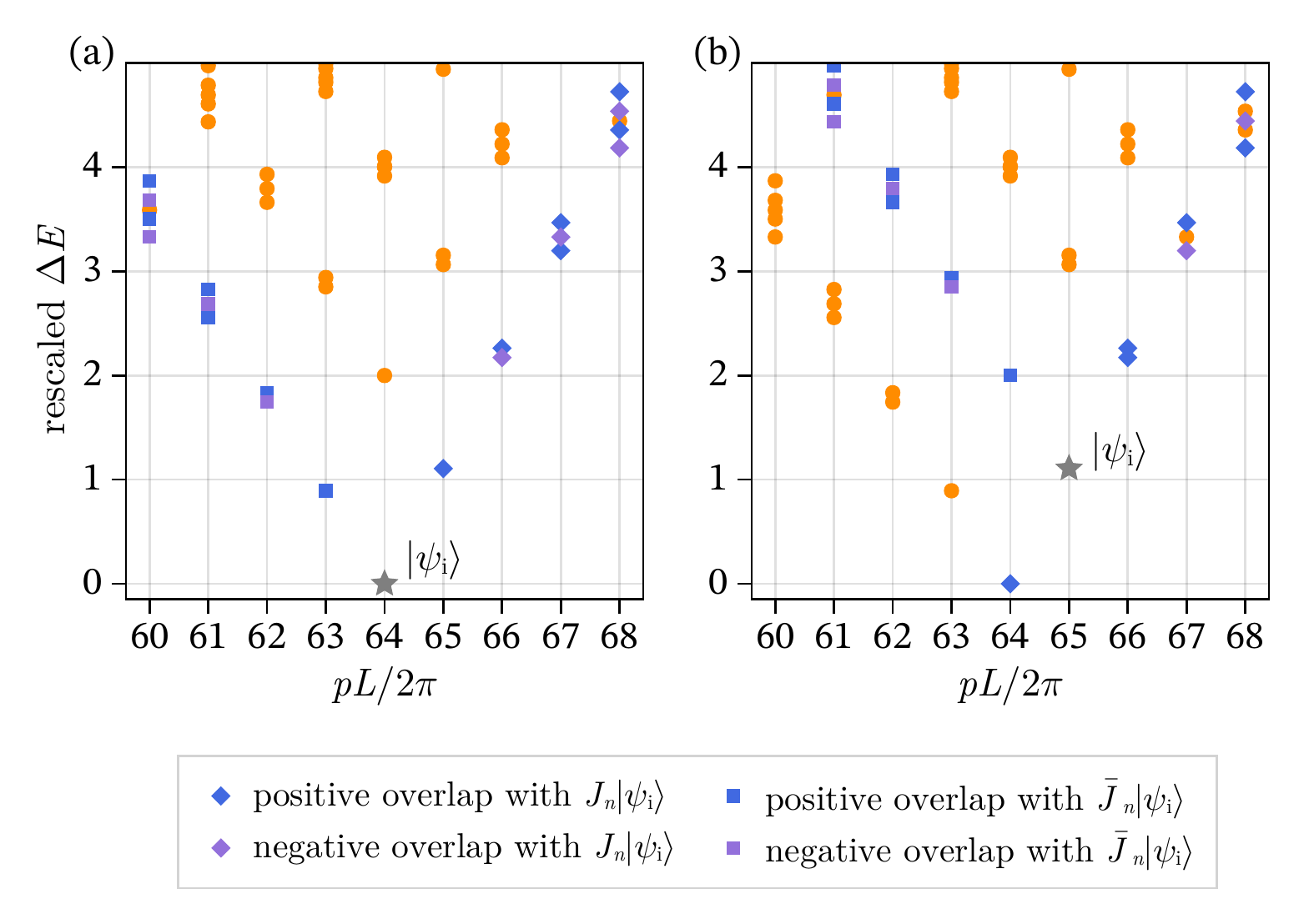}}
    \caption{Low energy spectrum near the state $|\psi_0^{m=1}\>$. The calculation is carried out in the Lieb-Liniger model with $c=1$, $L=64$ and particle number $N=64$. The states can be mapped from $|\psi_i\>$ (denoted by the gray star) with $J_n$'s and $\bar{J}_n$'s are marked with different shapes and colors. The initial states are chosen as (a) $|\psi_0^{m=1}\>$ and (b) $J_{-1} |\psi_0^{m=1}\>$. }
    \label{fig:ba-spect-branch-pi}
  \end{figure}

Finally, we look at the Kac-Moody tower at the momentum $p = 2\pi \rho_0$. 
The primary state in this tower correspond to the state with one cross-branch scattering [c.f. Fig.\ref{fig:fermi-sea} (c)], which we will refer to as $|\psi_0^{m=1}\>$.
As previously mentioned, the state $|\psi_0^{m=1}\>$ is a low-energy excitation state with a short length scale $\rho_0^{-1}$, which contributes to the $|m|=1$ term in Eq.~\eqref{eq:rho-theta-poisson-sum}.
When we apply the Kac-Moody generators to the state $|\psi_0^{m=1}\>$, the short-wavelength characteristics of the state will remain unchanged, as these generators can only produce particle-hole excitations that have long length scales.
This fact can be verified by calculating the form factors of the Kac-Moody generators.
In Fig.~\ref{fig:ba-spect-branch-pi}, we show the low-energy spectrum near $|\psi_0^{m=1}\>$. 
By calculating the form factors, we determine the states that can be obtained by acting with Kac-Moody generators on top of $|\psi_0^{m=1}\>$ and $J_{-1} |\psi_0^{m=1}\>$.
Comparing Fig.~\ref{fig:ba-spect} and Fig.~\ref{fig:ba-spect-branch-pi}, one can see that these two figures share almost the same feature, except that the spectrum in Fig.~\ref{fig:ba-spect-branch-pi} are slightly ``unbalanced'' due to the finite-size effect.

The form factors $C_{\mu, \lambda}[J_{n}]$ and $C_{\mu, \lambda}[\barJ_{n}]$ in the Lieb-Liniger model, along with their particle-hole excitation interpretation, are also discussed in Ref.~\cite{bouchoule-relaxation-2022}, which focuses on the case of the thermodynamic limit, in contrast with the finite-size study in our paper.

\section{Kac-Moody generators in continuous matrix product states} \label{sec:cmps}

For the numerical study of continuous systems, among others, the cMPS method~\cite{verstraete-continuous-2010,haegeman-calculus-2013} has become an indispensable technique. 
The cMPS method does not require the discretization of a continuous space and can therefore be directly applied to solve various ultracold atomic systems as well as (1+1)-dimensional quantum field theories~\cite{haegeman-applying-2010,rincon-liebliniger-2015,ganahl-continuous-2017-a,tuybens-variational-2022,lukin-continuous-2022}.
In addition to ground state simulations, cMPS can also be used to compute excited states~\cite{draxler-particles-2013} and time evolution~\cite{haegeman-quantum-2017,draxler-continuous-2017}.
In addition, cMPS can be related to continuous measurements~\cite{osborne-holographic-2010}, open quantum systems~\cite{kiukas-equivalence-2015}, classical stochastic dynamics~\cite{garrahan-classical-2016}, and thermodynamics of quantum lattice systems~\cite{tang-continuous-2020,tang-tensor-2021}.
Moreover, there also exist generalizations of the cMPS ansatz, such as the relativistic cMPS~\cite{tilloy-relativistic-2021}, and the continuous projected entangled-pair states~\cite{tilloy-continuous-2019,shachar-approximating-2022}. 

In this section, we will demonstrate that the behavior of the Kac-Moody generators  \eqref{eq:J-liebliniger} and \eqref{eq:barJ-liebliniger} can be correctly obtained from a cMPS simulation.

\paragraph{Ground state} To describe the ground state of the Lieb-Liniger model, we use a uniform, bosonic cMPS with periodic boundary conditions.
This circular bosonic cMPS is expressed as 
\begin{equation}
    | \Psi(Q, R) \> = \tr_{\mathrm{aux}} \left[ \calP \mathe^{\int_0^L \mathd x [Q\otimes \mathbbm{1} + R\otimes \hat{\psi}^\dagger(x)]} \right] |\Omega\>,
    \label{eq:cMPS}
\end{equation}
where $\mathcal{P}$ represents the path-ordering operator, $Q$ and $R$ are matrices of dimension $\chi \times \chi$ acting on the auxiliary space, $|\Omega\>$ is the Fock vacuum, and, $\mathbbm{1}$ and $\oppsi^{\dagger}(x)$ are respectively the identity operator and the boson creation operator acting on the physical space. 
The dimension $\chi$ of the auxiliary space is called the bond dimension of the cMPS. 
We obtain the ground state of a Hamiltonian $\hat{H}$ by minimizing the energy function variationally~\cite{verstraete-continuous-2010}
\begin{equation}
    E(Q, R) = \frac{\< \Psi(Q, R) |\hat{H}| \Psi(Q, R) \>}{\< \Psi(Q, R) | \Psi(Q, R) \>}.
    \label{eq:cmps_energy}
\end{equation}
Unlike the Bethe ansatz solution, in the cMPS simulation, the particle number cannot be fixed, and we need to introduce a chemical potential $\mu$ to regulate the number of particles in the system.

The details for evaluating the energy function \eqref{eq:cmps_energy} can be found in Appendix \ref{app:cMPS-computation-1}. 
The optimization of the cMPS can be performed using gradient-based optimization methods.
It is worth noting that the optimization of a circular cMPS is evidently more challenging than that of the infinite cMPS.
In our simulation, we employ Riemannian optimization
techniques~\cite{hauru-riemannian-2021}, the details of which are discussed in Appendix \ref{app:cMPS-optimization}.

\paragraph{Excited states} After obtaining the ground state represented as a uniform circular cMPS $|\Psi(Q, R)\>$, we can build low-excited states by introducing impurity matrices in the uniform cMPS~\cite{rommer-class-1997,pirvu-matrix-2012-b,haegeman-variational-2012,draxler-particles-2013,zou-conformal-2018}
\begin{align}
   |\Phi_p (V, W) \> = &\int_0^L \mathd x\, \mathe^{\mathi p x} \tr_{\mathrm{aux}} \left[ U(0, x)  \right. \nonumber \\
                         & \left. \times  [V \otimes \mathbbm{1} + W \otimes  \hat{\psi}^\dagger(x)] U(x, L)   \right] | \Omega \>,
    \label{eq:cMPS-excitation}
\end{align}
where $p$ is the momentum of the state, and $U(x, y) = \mathcal{P} \exp(\int_x^y dz [Q\otimes \mathbbm{1} + R \otimes \hat{\psi}^\dagger(z)])$. 
The $(\chi\times\chi)$-dimensional impurity matrices $V$ and $W$ introduce a single-particle excitation into the ground state, whose influence is within a range determined by the bond dimension of the cMPS. 
Although this excited-state ansatz is most suitable for single-particle excitations~\cite{draxler-particles-2013}, in principle, one can still obtain low-energy excited states accurately as long as the bond dimension $\chi$ is large enough~\footnote{As discussed in the Sec.~\ref{sec:liebliniger}, the low-energy excitations of interest in the Lieb-Liniger model are particle-hole excitations, which are classified as ``two-particle'' process in the current context, as it consists of two steps: creating a hole and then generating a particle.}.

To compute the excited states, we only needs to solve the following generalized eigenvalue problem~\cite{draxler-particles-2013}, 
\begin{equation}
    H_p V_{[V, W]} = E N_p V_{[V, W]}. \label{eq:generalized-eig}
\end{equation}
Here, $E$ is the energy of the excited state, and $V_{[V, W]}$ represents a $2\chi^2$-dimensional vector composed of the elements of $V$ and $W$.
$H_p$ and $N_p$ represents the effective Hamiltonian and the effective norm matrix in the space of $V_{[V, W]}$, which are defined by
\begin{align}
   \< \Phi_{p_1}(V_1, W_1) | \hat{H} | \Phi_{p_2}(V_2, W_2)\> = & \, L \, \delta(p_1, p_2)  \nonumber \\
        & \times V_{[V_1, W_1]}^{\dagger}  H_p V_{[V_2, W_2]}, \label{eq:effective-H} \\
   \< \Phi_{p_1}(V_1, W_1) | \Phi_{p_2}(V_2, W_2)\> = & \, L \, \delta(p_1, p_2)  \nonumber \\
        & \times V_{[V_1, W_1]}^{\dagger}  N_p V_{[V_2, W_2]}. \label{eq:effective-N}
\end{align}
The details for computing the matrix elements of $H_p$ and $N_p$ are included in Appendix \ref{app:cMPS-computation-2}. 

There are $\chi^2$ redundant degrees of freedom in the representation of $|\Phi_p(V, W)\>$, which can be traced back to the gauge redundancy in $|\Psi(Q, R)\>$~\cite{draxler-particles-2013}.
These gauge redundancies lead to zero eigenvalues in $\hat{H}$ and $\hat{N}$.
To fix this, we employ the following ``gauge-fixing'' condition
\begin{equation}
    \tr_{\bar{\mathcal{V}}} \left[ \mathe^{TL} (V \otimes \bar{I} + W \otimes \bar{R}) \right] = 0, 
    \label{eq:gauge-fixing-pbc}
\end{equation} 
where $\tr_{\bar{\mathcal{V}}}$ represents the partial trace over the auxiliary space $\bar{\mathcal{V}}$ where the matrices $\bar{I}$ and $\bar{R}$ live in, and $T = I \otimes \bar{Q} + Q \otimes \bar{I} + R \otimes \bar{R}$ is the cMPS transfer matrix.
This `gauge' condition also ensures that excited states with momentum zero, i.e., \ $|\Phi_0(V, W)\>$, have an exact zero physical overlap with the ground state, i.e., $\<\Psi(Q, R)|\Phi_0(V, W)\> = 0$. For all other momenta, orthogonality to the ground state is trivially ensured.

\paragraph{Form factors of Kac-Moody generators} With the cMPS approximations for the ground state and the low-energy excited states, it is straightforward to compute the form factors of the Kac-Moody generators.
The formulas for the computation of these form factors can be found in Appendix \ref{app:cMPS-computation-3}. 

Each (approximate) eigenstate obtained by solving the generalized eigenvalue problem \eqref{eq:generalized-eig} has an arbitrary phase, which will affect the phase of the form factors. This freedom can be used to ensure that the signs (but not the absolute value) of the form factors between a given initial state $|\psi_i\>$, e.g.\ the ground state, agree with the predictions of the bosonisation or Bethe ansatz approach. However, by then selecting a different initial state, namely one of the eigenstates whose phase is now fixed, and computing the form factors between this state and other eigenstates with a fixed phase, a nontrivial consistency check is obtained for the accuracy of our cMPS results.

\paragraph{Results}
Here, we present the results of the cMPS calculation.
The calculation is carried out in the Lieb-Liniger model with $c=1$, $\mu=1.426$, and $L=16$.
According to the Bethe ansatz solution, the ground state has the particle number $N_0 = 16$, and energy $E_{\mathrm{gs}} \approx -12.649511$.
The cMPS simulation is performed with bond dimension $\chi=20$.
The ground state calculation of the cMPS yields high precision, where the relative errors for the energy and particle number are $\epsilon_{E} \approx 2.5\times 10^{-5}$ and $\epsilon_{N} \approx 1.8 \times 10^{-6}$, respectively. 

\begin{figure}[!htb]
    \centering
    \resizebox{\columnwidth}{!}{\includegraphics{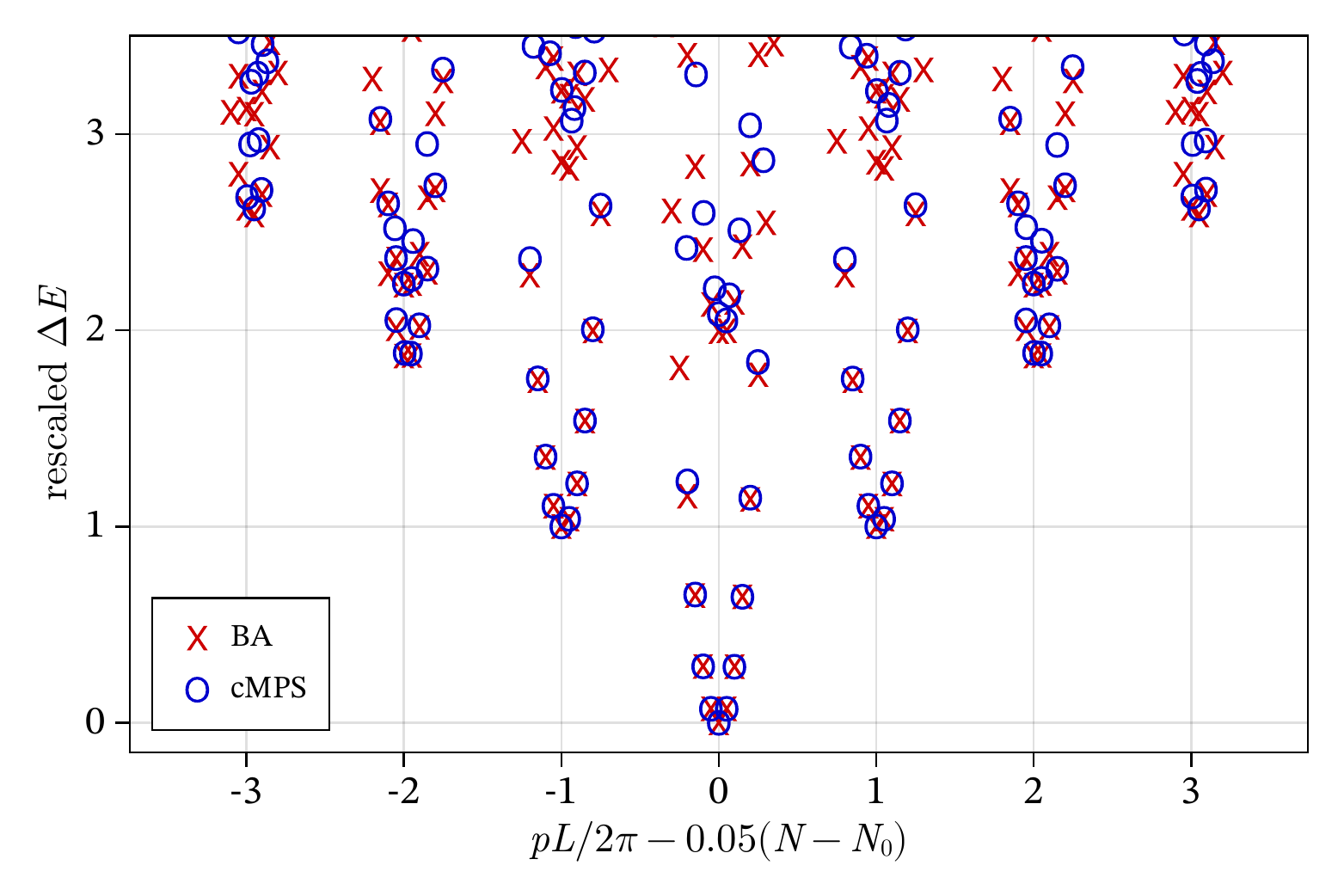}}
    \caption{The low-energy spectrum obtained from both Bethe ansatz (marked by crosses) and cMPS calculations (marked by circles). The energies of the states are rescaled. The horizontal axis is slightly shifted according to the numbers of particles in the states.}
    \label{fig:full-spect-cmps}
\end{figure}

Figure \ref{fig:full-spect-cmps} shows the low-energy excitation spectrum, which is obtained from both the Bethe ansatz solution and the cMPS calculations.
The horizontal axis is shifted slightly according to the number of particles in each state.
From Fig.~\ref{fig:full-spect-cmps}, it is clear that the cMPS calculation can correctly obtain the low-energy states, giving both energies and the particle numbers correctly, while it fails at higher energies. 

\begin{figure}[!htb]
    \centering
    \resizebox{\columnwidth}{!}{\includegraphics{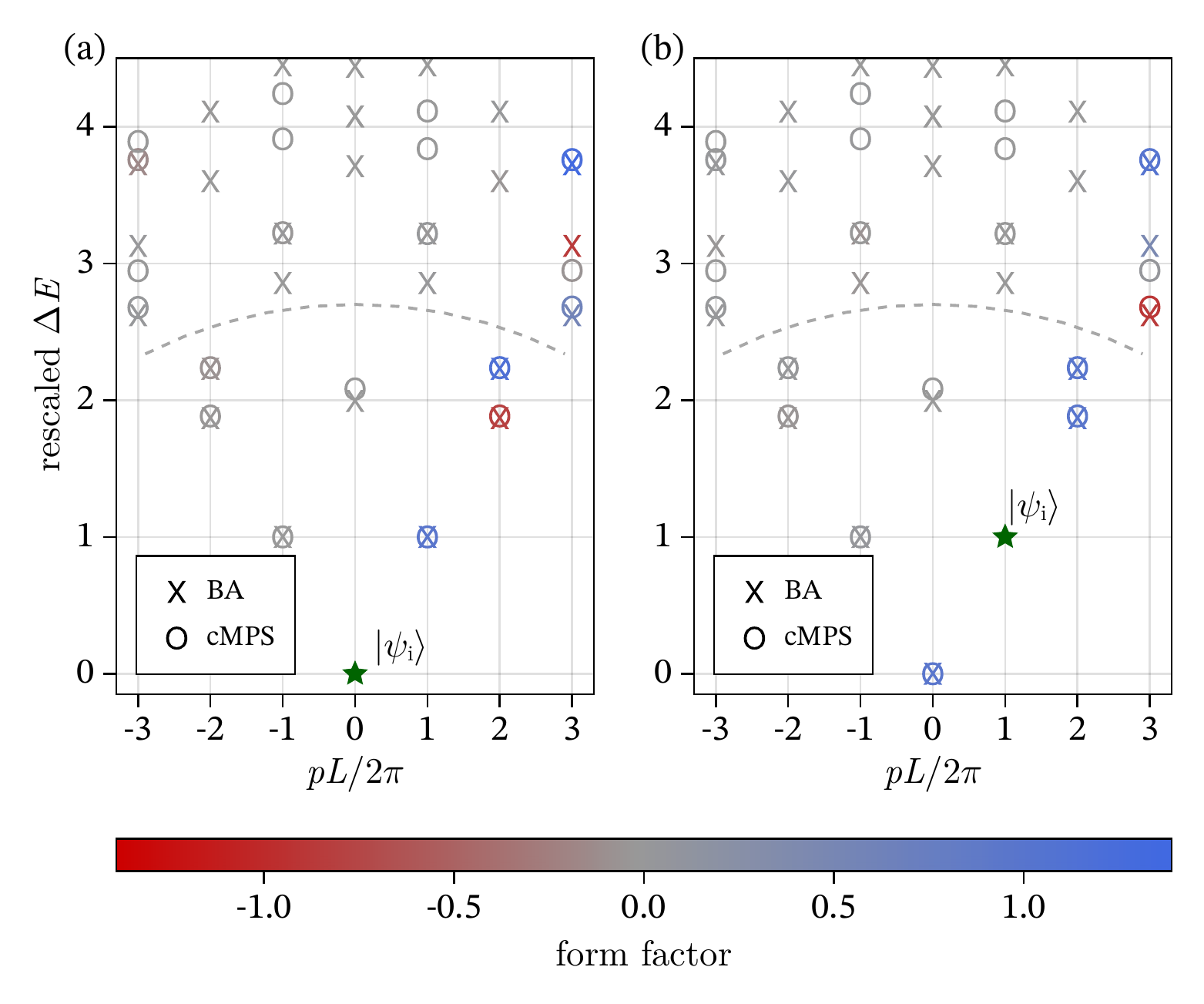}}
    \caption{The low-energy spectrum with fixed particle number $N=N_0$ and the results for the form factors. 
    In the spectrum, the Bethe ansatz results are marked as crosses and the cMPS results are marked as circles. 
    The values of the form factors are indicated by the colors of the data points, from which one can infer whether an eigenstate can be mapped from the initial state by mapping the Kac-Moody generators $J_{-n}$. 
    The initial states are chosen as (a) the ground state $|\psi_0 \>$ and (b) the excited state $J_{-1} |\psi_0\>$, which are marked by green stars. 
    For eigenstates under the gray dashed line, both the energy values and the form factors can be accurately obtained through the cMPS method.
    }
    \label{fig:Ngs-spect-cmps}
\end{figure}

In order to further demonstrate the effectiveness of the cMPS numerical calculation method, in Fig.~\ref{fig:Ngs-spect-cmps}, we compare the cMPS results of the form factors with the Bethe ansatz results,
where we choose two initial states, and then compute the form factor of $J_{-n}$ between the initial state and the other eigenstates. 
We only show the states with particle number $N=N_0$ in Fig.~\ref{fig:Ngs-spect-cmps} for the sake of clarity. 
Note that, for excited states obtained with cMPS, the selection of states with $N=N_0$ can only be done approximately, where we choose the states satisfying $|N - N_0| < 0.5$.
In Table \ref{tab:spectrum-data} we show both the norm and the phase of the form factors obtained by the cMPS method. 
From the Fig.~\ref{fig:Ngs-spect-cmps}, we can see that the form factors obtained from cMPS calculations are consistent with the Bethe ansatz solutions for the low-energy excited states, while they fail for eigenstates at higher energy levels.
We also list the detailed data for the lowest excited states in Table \ref{tab:spectrum-data}, which further demonstrates the quantitative accuracy of the cMPS results for these low-energy states.

\begin{table}[!htb]
\begin{tabular}{@{}llllllll@{}}
\toprule
\multirow{2}{*}{$\frac{\displaystyle pL}{\displaystyle 2\pi}$} & $N$ & \multicolumn{2}{l}{Scaled $\Delta E$} & \multicolumn{2}{l}{Form factors (a)} & \multicolumn{2}{l}{Form factors (b)}\\ \cmidrule(l){2-8} 
      & cMPS      & BA       & cMPS     & BA        & cMPS      & BA        & cMPS       \\ \midrule
$-2 $ & $15.8957$ & $1.8686$ & $1.8842$ & $-0.0266$ & $-0.0251$ & $-0.0345$ & $-0.0336$   \\
$-2$  & $15.9970$ & $2.2296$ & $2.2375$ & $-0.0638$ & $-0.0625$ & $0.0382$  & $0.0406$    \\
$-1$  & $16.0000$ & $1.0000$ & $1.0004$ & $0.0000$  & $-0.0004$ & $0.0523$  & $0.0509$    \\
$0$   & $15.9972$ & $1.9938$ & $2.0821$ & $0.0000$  & $0.0000$  & $0.0031$  & $0.0026$    \\
$1$   & $16.0000$ & $1.0000$ & $1.0004$ & $0.9992$  & $0.9993$  &   ---     & ---         \\
$2$   & $15.9556$ & $1.8686$ & $1.8844$ & $-0.7832$ & $-0.8009$ & $1.0321$  & $0.9764$    \\
$2$   & $15.9966$ & $2.2296$ & $2.2378$ & $1.1756$  & $1.1506$  & $0.9790$  & $1.0079$    \\ \bottomrule
\end{tabular}
\caption{Detailed data for the lowest seven excited states in the energy spectrum in Fig.~\ref{fig:Ngs-spect-cmps}.
We list the particle numbers, scaled energies and form factors of $J_{-n}$.
The form factors shown in both Fig.~\ref{fig:Ngs-spect-cmps} (a) and (b) are listed.
The results listed are obtained by Bethe ansatz (BA) and cMPS calculations, respectively, except that for the particle numbers we only list the cMPS results, since the particle number in Bethe ansatz solution is exact.
All numbers in the table are accurate to 4 decimal places.}
\label{tab:spectrum-data}
\end{table}

We note that, although the $\mathrm{U}(1)$ symmetry is not implemented in the cMPS ansatz, the cMPS is still able to obtain the correct particle number to a high precision for both the ground state and the low-energy excited states.
Moreover, from the form factor results, we can also infer that the structure of the particle hole excitation is also effectively encoded in the excited states obtained with cMPS method.

On the other hand, due to the lack of $\mathrm{U}(1)$ symmetry in the cMPS ansatz, our cMPS results can only cover a part of the Kac-Moody tower, since the numbers of particles in most of the low-energy states are different from that of the ground state. 
This also makes it difficult to push the cMPS calculation to larger systems, where it becomes more difficult to calculate the excited states accurately. 

\section{Conclusion and Outlook} \label{sec:conclusion-and-outlook}
In summary, we have extended the lattice realization of the Kac-Moody generators to continuous systems and applied it to one-dimensional continuous bosonic systems.
We have justified this microscopic realization of Kac-Moody generators in two different ways: by phenomenological bosonization and by studying the integrable Lieb-Liniger model. 
We have also tested the computation of the Kac-Moody generator in the cMPS simulations, which can be used for more challenging problems where there are no exact solutions.

The Kac-Moody generator can be interpreted as describing particle-hole excitations in a particular fermionic picture.
In the integrable Lieb-Liniger model, we have shown that this fermionic picture corresponds to the distribution of the quantum numbers in the Bethe wavefunctions. 
It would be interesting to further investigate other integrable systems to check the generality of this result, such as the Calogero-Sutherland model~\cite{calogero-ground-1969,sutherland-quantum-1971-a,sutherland-quantum-1971-b}, the Haldane-Shastry model~\cite{haldane-exact-1988,shastry-exact-1988}, and the XXZ model~\cite{gaudin-bethe-book-2014}.
Another possible direction for further study is to analyze the effect of different boundary conditions, such as open and twisted boundary conditions.
Furthermore, as the Bethe wavefunction in the Lieb-Liniger model has an exact cMPS representation~\cite{maruyama-continuous-2010}, it would be worthwhile to explore its potential utility in evaluating the form factors.

For non-integrable systems, one usually has to resort to numerical simulation methods.
An efficient numerical method to compute form factors for Kac-Moody generators can also complement the form factor techniques in the Bethe ansatz, since the form factors for many integrable systems are very difficult to compute analytically. 
In our paper, we have tested the Kac-Moody generator realization in the cMPS simulation of the Lieb-Liniger model.
A natural direction for further study is to simulate non-integrable systems with cMPS, such interacting boson systems with long-range interactions~\cite{rincon-liebliniger-2015}. Note that the Kac-Moody generators constructed in our paper do not depend on the microscopic details in the Hamiltonian, but only on the symmetry of the system. 
This makes it possible to construct the Kac-Moody generator for systems where the Hamiltonian is not available.
One example is the cMPS in the continuous matrix product operator simulation~\cite{tang-continuous-2020,tang-tensor-2021}, where the cMPS is the dominant eigenvector of the quantum transfer matrix. Another possible research direction is to consider possible extensions of the cMPS ansatz with a fixed particle number. This could help us to exclude states with different particle numbers in the low-energy spectrum and thus study the states in a single Kac-Moody tower more efficiently. Moreover, this also allows us to examine how this symmetry would be encoded in the Kac-Moody generator. While such extensions can be constructed, it is currently unclear whether they can be efficiently optimized.

Finally, we remark that the techniques developed in our paper can be generalized and applied to other continuous systems with $\mathrm{SU}(2)_k$ and other Kac-Moody algebras, such as multicomponent boson systems, spin-$1/2$ fermionic systems, and multicomponent Sutherland models.

\paragraph*{Note.} Our code implementations and data are available at \footnote{See \url{https://github.com/tangwei94/LiebLinigerBA.jl} and \url{https://github.com/tangwei94/CircularCMPS.jl} for implementations for the Bethe ansatz solution and the cMPS calculations. 
The scripts and raw data for the results shown in the paper can be found at \url{https://github.com/tangwei94/LiebLinigerKacMoody}.}.

\section*{Acknowledgment} 
We thank Hong-Hao Tu and Jacopo De Nardis for helpful discussions. 
This paper received funding from the European Research Council (ERC) under the European Unions Horizon 2020 research and innovation programme (Grant Agreement No 715861 (ERQUAF)).

\appendix

\onecolumngrid

\section{Form factors of density and current operator in Bethe ansatz}\label{app:form-factors-ba}
In this appendix, we give the form factors of the density operator $n(0)$ and the density current operator $j(0)$, which are calculated using the algebraic Bethe ansatz approach~\cite{slavnov-nonequal-1990,de-nardis-density-2015}.
For Bethe states $|\thesetof{\mu_j}\rangle$ and $\thesetof{\lambda_j}\rangle$, we have
\begin{align}
    \<\thesetof{\mu_j}| n(0) |\thesetof{\lambda_j}\> = & \left(\sum_{j=1}^N (\mu_j - \lambda_j)\right) 
    \prod_{j=1}^N (V_j^+ - V_j^-) \times \prod_{j,k}^N \left(\frac{\lambda_j - \lambda_k + \mathi c}{\mu_j - \lambda_k}\right) 
    \frac{\det(\delta_{j k} + U_{j k})}{V_p^+ - V_p^-}, \label{eq:form_factor_rho}\\
    \<\thesetof{\mu_j}| j(0) |\thesetof{\lambda_j}\> = & \left(\sum_{j=1}^N (\mu_j^2 - \lambda_j^2)\right) 
    \prod_{j=1}^N (V_j^+ - V_j^-) \times \prod_{j,k}^N \left(\frac{\lambda_j - \lambda_k + \mathi c}{\mu_j - \lambda_k}\right) 
    \frac{\det(\delta_{j k} + U_{j k})}{V_p^+ - V_p^-}, \label{eq:form_factor_p}\\
    \< \thesetof{\lambda_j} | \thesetof{\lambda_j} \> = & \, c^N \prod_{j \neq k}^N \frac{\lambda_j - \lambda_k + \mathi c}{\lambda_j - \lambda_k} \det \mathcal{G}, \label{eq:gaudin_norm}
\end{align}
where 
\begin{align}
    \mathcal{G}_{jk} &= \delta_{jk} (L + \sum_{m=1}^N K(\lambda_j - \lambda_m)) - K(\lambda_j - \lambda_k), \\
    K(\lambda) &= 2c/(\lambda^2 + c^2), \label{eq:K_lambda}\\
    V_j^{\pm} &= \prod_{k=1}^N \frac{\mu_k - \lambda_j \pm \mathi c}{\lambda_k - \lambda_j \pm \mathi c}, \\
    U_{jk} &= \mathi \frac{\mu_j-\lambda_j}{V_j^+ - V_j^-} \prod_{m\neq j}^N \left( \frac{\mu_m - \lambda_j}{\lambda_m - \lambda_j} \right) (K(\lambda_j - \lambda_k) - K(\lambda_p - \lambda_k)).
\end{align}

\section{cMPS computation details} \label{app:cMPS-computation}

In this appendix we include the details for evaluating the cMPS formulas, which are carried out using the techniques in Ref.~\cite{haegeman-calculus-2013}. 

\subsection{Computation of the energy function with circular cMPS} \label{app:cMPS-computation-1}

To compute Eq.~\eqref{eq:cmps_energy}, we first recall the following relations 
\begin{align}
    \psi(x) |\Psi (Q, R) \> &= \tr_{\mathrm{aux}} \left[ U(0, x) (R \otimes \mathbbm{1}) U(x, L) \right], \\
    \psi(x)\psi(x) |\Psi (Q, R) \> &= \tr_{\mathrm{aux}} \left[ U(0, x) (R^2 \otimes \mathbbm{1}) U(x, L) \right], \\
    \partial_x \psi(x) |\Psi (Q, R) \> &= \tr_{\mathrm{aux}} \left[ U(0, x) ([Q, R] \otimes \mathbbm{1}) U(x, L) \right], 
\end{align}
where $U(x, y) = \mathcal{P} \exp(\int_x^y dz [Q\otimes \mathbbm{1} + R \otimes \hat{\psi}^\dagger(z)])$.
From these relations, we can obtain
\begin{align}
    \< \psi^\dagger(x) \psi(x) \> &= \frac{1}{\mathcal{N}} \tr_{\mathrm{aux}} \left[ \mathe^{T L} (R \otimes \bar{R}) \right], \label{eq:ob-cmps-chem}\\
    \< \psi^\dagger(x)\psi^\dagger(x) \psi(x)\psi(x) \> &= \frac{1}{\mathcal{N}} \tr_{\mathrm{aux}} \left[ \mathe^{T L} (R^2 \otimes \bar{R}^2) \right], \label{eq:ob-cmps-interaction}\\
    \left\langle \partial_x\psi^\dagger(x)\partial_x\psi(x) \right\rangle &= \frac{1}{\mathcal{N}} \tr_{\mathrm{aux}} \left[ \mathe^{T L} ([Q, R] \otimes [\bar{Q}, \bar{R}]) \right],\label{eq:ob-cmps-kinetic}
\end{align}
where $T = I \otimes \bar{Q} + Q \otimes \bar{I} + R \otimes \bar{R}$ is the cMPS transfer matrix, and $\mathcal{N} = \tr_{\mathrm{aux}}[\exp(TL)]$ is the squared norm of the cMPS. 
Using Eqs.~\eqref{eq:ob-cmps-chem}, \eqref{eq:ob-cmps-interaction} and \eqref{eq:ob-cmps-kinetic}, one can easily evaluate the energy function \eqref{eq:cmps_energy}.  

\subsection{Computation of the effective Hamiltonian and the effective norm matrix} \label{app:cMPS-computation-2}

To compute the effective Hamiltonian $H_p$ in Eq.~\eqref{eq:effective-H}, we first compute
\begin{align}
    \psi(x) |\Phi_p (V, W) \> =& \int_x^{x+L} \mathd y \,\mathe^{\mathi p y} \tr_{\mathrm{aux}} \left[ (R \otimes \mathbbm{1}) U(x, y) (V \otimes \mathbbm{1} + W \otimes \psi^\dagger(y)) U(y, x+L)  \right]  \nonumber \\
    & + \mathe^{\mathi p x} \tr_{\mathrm{aux}} \left[ U(0, x) (W \otimes \mathbbm{1}) U(x, L) \right], \\
    \psi(x)\psi(x) |\Phi_p (V, W) \> =& \int_x^{x+L} \mathd y \,\mathe^{\mathi p y} \tr_{\mathrm{aux}} \left[ (R^2 \otimes \mathbbm{1}) U(x, y) (V \otimes \mathbbm{1} + W \otimes \psi^\dagger(y)) U(y, x+L)  \right]  \nonumber \\
    & + \mathe^{\mathi p x} \tr_{\mathrm{aux}} \left[ U(0, x) ((R W + W R) \otimes \mathbbm{1}) U(x, L) \right], \\
    \partial_x \psi(x) |\Phi_p (V, W) \> =& \int_x^{x+L} \mathd y \,\mathe^{\mathi p y} \tr_{\mathrm{aux}} \left[ ([Q, R] \otimes \mathbbm{1}) U(x, y) (V \otimes \mathbbm{1} + W \otimes \psi^\dagger(y)) U(y, x+L)  \right]  \nonumber \\
    & + \mathe^{\mathi p x} \tr_{\mathrm{aux}} \left[ U(0, x) (([V, R] + [Q, W] + \mathi p W) \otimes \mathbbm{1}) U(x, L) \right].
\end{align}

Before we proceed to compute the overlaps in Eqs.~\eqref{eq:cmps_energy}, we first introduce the following notations.
Suppose the cMPS transfer matrix $T$ has the eigendecomposition $T = U \Lambda U^{-1}$. For $\chi^2\times\chi^2$ matrices $A$, $B$, and $C$ and momentum $p_{AB}$ and $p_{BC}$, we define
\begin{align}
    \mathcal{C}_2 (p_{AB} \mid A, B) &\equiv  
    \int_x^{x+L} \mathd y 
    \,\tr_{\mathrm{aux}} \left[ 
        A \mathe^{(y - x) (T + \mathi p_{AB})} 
        B \mathe^{(x - y + L) T} \right] \label{eq:C2-integral} \\
    &= \sum_{s k} \theta_2 (\Lambda_s + \mathi p_{AB}, \Lambda_k) (U^{-1} A U)_{sk} (U^{-1} B U)_{ks}, \\
    \mathcal{C}_3 (p_{AB}, p_{BC}\mid A, B, C) &\equiv 
    \int_x^{x+L} \mathd y' 
    \int_{y'}^{x+L} \mathd y
    \, \tr_{\mathrm{aux}} \left[ 
        A \mathe^{(y' - x) (T + \mathi p_{AB})} 
        B \mathe^{(y - y') (T + \mathi p_{BC})} 
        C \mathe^{(x - y + L) T} \right] \label{eq:C3-integral} \\
    & = \sum_{s k l} \theta_3 (\Lambda_k + \mathi p_{AB}, \Lambda_l + \mathi p_{BC}, \Lambda_s) 
    (U^{-1} A U)_{s k} 
    (U^{-1} B U)_{k l} 
    (U^{-1} C U)_{l s}, 
\end{align}
where  
\begin{align}
    \theta_2(a, b) &= \frac{\mathe^{La} - \mathe^{Lb}}{a - b},&
    \theta_3 (a, b, c) &= \frac{a(\mathe^{Lb} - \mathe^{Lc}) + b(\mathe^{Lc} - \mathe^{La}) + c(\mathe^{La} - \mathe^{Lb})}{(a-b)(b-c)(c-a)}.
\end{align}
Here, we have evaluated the integrals in Eqs.~\eqref{eq:C2-integral} and \eqref{eq:C3-integral} analytically. 
Alternatively, one can also use the Gaussian quadrature to evaluate the integrals numerically~\cite{press-numerical-book-2007}, which allows highly efficient parallelization. A final method, which we will explore elsewhere, is to exploit that the exponential of an upper block triangular matrix is given by
\begin{equation}
    \exp\left(L \begin{bmatrix} T + \mathi p_{AB} & B \\ 0 & T\end{bmatrix}\right) = \begin{bmatrix} \mathe^{L(T+ \mathi p_{AB})} & \int_0^L \mathd y\,  \mathe^{x (T+ \mathi p_{AB})} B \mathe^{(L-x) T} \\ 0  & \mathe^{L T}\end{bmatrix}, 
\end{equation}
where the matrix exponential can be computed using the Pad\'e approximation. The upper right block in the right hand side can then be multiplied with $A$ and traced over to yield $\mathcal{C}_2 (p_{AB} \mid A, B)$. A similar approach can be used for $\mathcal{C}_3 (p_{AB}, p_{BC} \mid A, B, C)$ by using a $3 \times 3$ block matrix.

Combining the equations above, we proceed to compute the matrix elements of different Hamiltonian terms separately.
For the particle density, we have
\begin{align}
    &\int_0^L \mathd x \< \Phi_{p'} (V', W') | \psi^\dagger(x) \psi(x) | \Phi_{p} (V, W) \>  \nonumber\\
    = &\, L \delta_{p, p'} \left[\mathcal{C}_3\left(p-p', p \mid R\otimes\bar{R}, I\otimes\bar{V}' + R\otimes\bar{W}', V\otimes\bar{I} + W\otimes\bar{R}\right) + \right.\nonumber\\
     &\, \hphantom{L \delta_{p, p'} [} \mathcal{C}_3\left(p-p', -p' \mid R\otimes\bar{R}, V\otimes\bar{I} + W\otimes\bar{R}, I\otimes\bar{V}' + R\otimes\bar{W}'\right) + \nonumber\\ 
     &\, \hphantom{L \delta_{p, p'} [} \mathcal{C}_2\left(p-p' \mid R\otimes\bar{R} , W\otimes \bar{W}' \right) + 
     \mathcal{C}_2\left(-p' \mid W\otimes\bar{R} , I\otimes\bar{V}' + R\otimes\bar{W}' \right) + \nonumber\\
     &\hphantom{L \delta_{p, p'} [} \left. \mathcal{C}_2\left(p \mid R\otimes\bar{W}' , V\otimes\bar{I} + W\otimes\bar{R} \right) +
     \tr_{\mathrm{aux}} \left(\mathe^{L T} (W\otimes\bar{W}') \right) \right].\label{eq:ob-excitation-chem}
\end{align}
For the kinetic energy, we have
\begin{align}
    &\int_0^L \mathd x \< \Phi_{p'} (V', W') |  \partial_x\psi^\dagger(x)\partial_x\psi(x)| \Phi_{p} (V, W) \>  \nonumber\\
    = & \,L \delta_{p, p'}  \left[ \mathcal{C}_3\left(p-p', p \mid [Q, R]\otimes[\bar{Q}, \bar{R}], I\otimes\bar{V}' + R\otimes\bar{W}', V\otimes\bar{I} + W\otimes\bar{R}\right) +\right. \nonumber\\
     & \, \hphantom{L \delta_{p, p'} [} \mathcal{C}_3\left(p-p', -p' \mid [Q, R]\otimes[\bar{Q}, \bar{R}], V\otimes\bar{I} + W\otimes\bar{R}, I\otimes\bar{V}' + R\otimes\bar{W}'\right) + \nonumber\\ 
     & \, \hphantom{L \delta_{p, p'} [} \mathcal{C}_2\left(p-p' \mid [Q, R]\otimes[\bar{Q}, \bar{R}], W\otimes \bar{W}' \right) + 
     \mathcal{C}_2\left(-p' \mid K \otimes[\bar{Q}, \bar{R}], I\otimes\bar{V}' + R\otimes\bar{W}' \right) + \nonumber\\
     &  \hphantom{L \delta_{p, p'} [} \left. \mathcal{C}_2\left(p \mid [Q, R]\otimes \bar{K}', V\otimes\bar{I} + W\otimes\bar{R} \right) +
     \tr_{\mathrm{aux}} \left(\mathe^{L T} (K \otimes\bar{K}') \right) \right], \label{eq:ob-excitation-kin}
\end{align}
where $K = [V, R] + [Q, W] + \mathi p W$ and $K' = [V', R] + [Q, W'] + \mathi p' W'$.
For the interaction term, we have 
\begin{align}
    &\int_0^L \mathd x \< \Phi_{p'} (V', W') | \psi^\dagger(x)\psi^\dagger(x)\psi(x)\psi(x) | \Phi_{p} (V, W) \>  \nonumber\\
    = & \,L \delta_{p, p'}  \left[ \mathcal{C}_3\left(p-p', p \mid R^2\otimes\bar{R}^2, I\otimes\bar{V}' + R\otimes\bar{W}', V\otimes\bar{I} + W\otimes\bar{R}\right) + \right.\nonumber\\
     & \, \hphantom{L \delta_{p, p'} [} \mathcal{C}_3\left(p-p', -p' \mid R^2\otimes\bar{R}^2, V\otimes\bar{I} + W\otimes\bar{R}, I\otimes\bar{V}' + R\otimes\bar{W}'\right) + \nonumber\\ 
     & \, \hphantom{L \delta_{p, p'} [} \mathcal{C}_2\left(p-p' \mid R^2\otimes\bar{R}^2, W\otimes \bar{W}' \right) + 
     \mathcal{C}_2\left(-p' \mid(RW + WR) \otimes \bar{R}^2, I\otimes\bar{V}' + R\otimes\bar{W}' \right) + \nonumber\\
     & \hphantom{L \delta_{p, p'} [} \left. \mathcal{C}_2\left(p \mid R^2 \otimes (\bar{R}\bar{W}' + \bar{W}'\bar{R}), V\otimes\bar{I} + W\otimes\bar{R} \right) +
     \tr_{\mathrm{aux}} \left(\mathe^{L T} ((RW + WR) \otimes(\bar{R} \bar{W}' + \bar{W}' \bar{R})) \right)\right].\label{eq:ob-excitation-interaction}
\end{align}
Combining Eqs.~\eqref{eq:ob-cmps-chem}, \eqref{eq:ob-cmps-interaction} and \eqref{eq:ob-cmps-kinetic}, we can compute the matrix elements for $H_p$. 

To compute the effective norm matrix $N_p$ in Eq.~\eqref{eq:effective-N}, we have
\begin{equation}
    \< \Phi_{p'}(V', W')| \Phi_p(V, W)\> = \delta(p-p') \left[ \tr_{\mathrm{aux}} [\mathe^{L T} (W\otimes \bar{W}')] + \mathcal{C}_2\left(-p \mid V\otimes\bar{I} + W\otimes\bar{R}, I\otimes\bar{V}'+R\otimes\bar{W}'\right) \right].
\end{equation}

\subsection{Computation of the form factors of the Kac-Moody generators} \label{app:cMPS-computation-3}

In the framework of cMPS, the form factors of the Kac-Moody generators \eqref{eq:J-liebliniger} and \eqref{eq:barJ-liebliniger} are calculated in two scenarios.
First, to compute the form factor between the ground state and the excited states, we need to calculate 
\begin{align}
    \int_0^L \mathe^{\mathi q x} \mathd x \< \Phi_{p} (V, W) | \rho(x) | \Psi(Q, R) \> 
    &= L \delta_{p, q} \left[ 
        \mathcal{C}_2\left(-p \mid R \otimes \bar{R}, I\otimes\bar{V} + R\otimes\bar{W} \right) +
     \tr_{\mathrm{aux}} \left(\mathe^{L T} (R \otimes \bar{W})  \right)\right],\\
    \int_0^L \mathe^{\mathi q x} \mathd x \< \Phi_{p} (V, W) | j(x) | \Psi(Q, R) \> 
    &= \mathi L \delta_{p, q} \left[ 
        \mathcal{C}_2\left(-p \mid [Q, R] \otimes \bar{R} - R \otimes [\bar{Q}, \bar{R}], I\otimes\bar{V} + R\otimes\bar{W} \right) + \right. \nonumber \\
       &\hspace{1.75cm} \left. \tr_{\mathrm{aux}} \left(\mathe^{L T} (R \otimes \bar{K} - [Q, R] \otimes \bar{W})  \right)\right],
\end{align}
where $K = [V, R] + [Q, W] + \mathi p W$.
Second, to compute the form factor between excited states, we have
\begin{align}
    &\int_0^L \mathe^{\mathi q x} \mathd x \< \Phi_{p'} (V', W') | \rho(x) | \Phi_{p} (V, W) \>  \nonumber\\
    = &\, L \delta_{p+q, p'} \left[\mathcal{C}_3\left(p-p', p \mid R\otimes\bar{R}, I\otimes\bar{V}' + R\otimes\bar{W}', V\otimes\bar{I} + W\otimes\bar{R}\right) + \right.\nonumber\\
     &\, \hphantom{L \delta_{p+q, p'} [} \mathcal{C}_3\left(p-p', -p' \mid R\otimes\bar{R}, V\otimes\bar{I} + W\otimes\bar{R}, I\otimes\bar{V}' + R\otimes\bar{W}'\right) + \nonumber\\ 
     &\, \hphantom{L \delta_{p+q, p'} [} \mathcal{C}_2\left(p-p' \mid R\otimes\bar{R} , W\otimes \bar{W}' \right) + 
     \mathcal{C}_2\left(-p' \mid W\otimes\bar{R} , I\otimes\bar{V}' + R\otimes\bar{W}' \right) + \nonumber\\
     &\hphantom{L \delta_{p+q, p'} [} \left. \mathcal{C}_2\left(p \mid R\otimes\bar{W}' , V\otimes\bar{I} + W\otimes\bar{R} \right) +
     \tr_{\mathrm{aux}} \left(\mathe^{L T} (W\otimes\bar{W}') \right) \right].\label{eq:formfactor-excitation-rho} \\
    &\int_0^L \mathe^{\mathi q x} \mathd x \< \Phi_{p'} (V', W') | j(x) | \Phi_{p} (V, W) \>  \nonumber\\
    = &\, \mathi L \delta_{p+q, p'} \left[\mathcal{C}_3\left(p-p', p \mid [Q, R] \otimes\bar{R} - R \otimes [\bar{Q}, \bar{R}], I\otimes\bar{V}' + R\otimes\bar{W}', V\otimes\bar{I} + W\otimes\bar{R}\right) + \right.\nonumber\\
     &\, \hphantom{\mathi L \delta_{p+q, p'} [} \mathcal{C}_3\left(p-p', -p' \mid [Q, R] \otimes\bar{R} - R \otimes [\bar{Q}, \bar{R}], V\otimes\bar{I} + W\otimes\bar{R}, I\otimes\bar{V}' + R\otimes\bar{W}'\right) + \nonumber\\ 
     &\, \hphantom{\mathi L \delta_{p+q, p'} [} \mathcal{C}_2\left(p-p' \mid [Q, R] \otimes\bar{R} - R \otimes [\bar{Q}, \bar{R}] , W\otimes \bar{W}' \right) + 
     \mathcal{C}_2\left(-p' \mid K\otimes\bar{R} - W \otimes [\bar{Q}, \bar{R}] , I\otimes\bar{V}' + R\otimes\bar{W}' \right) + \nonumber\\
     &\hphantom{\mathi L \delta_{p+q, p'} [} \left. \mathcal{C}_2\left(p \mid [Q, R] \otimes \bar{W}' - R\otimes\bar{K}' , V\otimes\bar{I} + W\otimes\bar{R} \right) +
     \tr_{\mathrm{aux}} \left(\mathe^{L T} (K\otimes\bar{W}' - W\otimes \bar{K}') \right) \right],\label{eq:formfactor-excitation-j}
\end{align}
where $K = [V, R] + [Q, W] + \mathi p W$ and $K' = [V', R] + [Q, W'] + \mathi p' W'$.

\twocolumngrid
\section{Riemannian optimization of the circular cMPS} \label{app:cMPS-optimization}

As mentioned in Sec.~\ref{sec:cmps}, we represent the ground state as a circular uniform cMPS $|\Psi(Q, R)\>$, and then minimize the energy function $E(Q, R)$ using gradient-based optimization.
The derivative $\partial_{(Q,R)}E(Q, R)$ can be obtained by manually working out its expression, or by using the automatic differentiation framework~\cite{liao-differentiable-2019}.

With the energy function and its derivative, the most straightforward way to implement the optimization is to use standard optimization algorithms, such as L-BFGS and conjugate-gradient descent. 
However, for the circular uniform cMPS, the optimization problem is highly nonlinear. 
A typical scenario in such standard optimization of the circular cMPS is that the optimization requires a large number of optimization steps, which quickly becomes formidable as one increases the bond dimension.
Among others, one major reason for the difficulty in the optimization comes from the conditioning of the Hessian, which remains nearly singular even after the gauge redundancies in the cMPS are eliminated. 
This nearly singular Hessian matrix makes the landscape of the energy function highly irregular and greatly slows down the optimization procedure.
For example, in quasi-Newton algorithms like L-BFGS, the algorithm maintains an approximation for the Hessian and use it to determine the search direction, and the nearly singular property of the Hessian severely impedes this approximation process. 
Such singular Hessian matrices are not only seen in cases of circular uniform cMPS optimizations~\footnote{For other examples, see e.g., Refs.~\cite{hauru-riemannian-2021,tilloy-relativistic-2021}}, and one may expect this is a common difficulty faced by the straightforward gradient optimization of tensor network wavefunctions.

In this paper, we employ the Riemannian optimization method for isometric tensor networks~\cite{hauru-riemannian-2021}. 
By restricting the cMPS to the left-canonical form, we restrict the cMPS local tensor to the Grassmann manifold (or some particular limit thereof), and use the Riemannian generalization of the L-BFGS algorithm to optimize the cMPS. 
Furthermore, we construct a preconditioner for the L-BFGS algorithm, which can largely mitigate the difficulties mentioned above. 

\paragraph{Grassmann manifold} The cMPS can be obtained as the limit $\epsilon\to 0$ of a MPS in which the local tensor $A$ takes the particular form 
\begin{equation}
    A = \binom{\mathbbm{1} + \epsilon Q}{\sqrt{\epsilon} R}.
\end{equation}
In the left-canonical form, the local tensor $A$ is an isometric tensor which can be taken to live in the Grassmann manifold, because of the remaining unitary gauge freedom that remains in the left-canonical MPS format. In the $\epsilon\to 0$ limit, the isometry condition translates into the requirement that the matrices $Q$ and $R$ should satisfy
\begin{equation}
    Q + Q^\dagger + R^\dagger R = 0.
    \label{eq:left-canonical-condition}
\end{equation}
In the following discussions, we will perform the optimization within the manifold formed by the matrices $(Q, R)$ satisfying Eq.~\eqref{eq:left-canonical-condition} and refer to it as the Grassmann manifold.

\paragraph{Gradient and search direction} In the Grassmann manifold, one can attach a tangent space at each point $(Q, R)$. In our case, the tangent vector $(V, W)$ should satisfy
\begin{equation}
    V = \mathrm{i} K - R^\dagger W,
    \label{eq:tangent-vector-one} 
\end{equation}
where $K$ is a Hermitian matrix, so that $(Q+\alpha V, R + \alpha W)$ still satisfy the condition \eqref{eq:left-canonical-condition} to the first order of $\alpha$. However, the parameters in $K$ correspond exactly to the remaining unitary gauge freedom, and a physically equivalent tangent vector can be obtained with the simpler parametrization
\begin{equation}
    V = - R^\dagger W.
    \label{eq:tangent-vector} 
\end{equation}
Equation \eqref{eq:tangent-vector} allows us to parametrize the tangent vector solely by the matrix $W$ and define an inner product between the tangent vectors as 
\begin{equation}
    \< (V_1, W_1), (V_2, W_2)\> = \tr(W_1^\dagger W_2).
    \label{eq:tangent-vector-dot}
\end{equation}
One should not confuse this inner product with the physical overlap $\< \Phi(V_1, W_1) |\Phi(V_2, W_2) \>$ between tangent vectors. The latter would give rise to a more complicated inner product in terms of the $W_1$ and $W_2$ parameters. Within the context of the Riemannian optimization methods, we prefer to work with the simpler (but unphysical) Euclidean inner product in Eq.~\eqref{eq:tangent-vector-dot}.

At each optimization step, we will first project the derivative $(\bar{Q}, \bar{R})$ into the tangent space, the result of which will henceforth be referred to as the \emph{gradient} at $(Q, R)$. 
The \emph{search direction} is then determined by the L-BFGS algorithm based on the gradients in the current and previous optimization steps.
During the optimization, we will always restrict the gradient and the search direction to the tangent space.

To determine the gradient from the partial derivative $(\bar{Q}, \bar{R})$, consider a random tangent vector $(-R^\dagger W, W)$.
We note that the infinitesimal change of the energy function along this vector should be given by the inner product between $(-R^\dagger W, W)$ and the gradient $(-R^\dagger W_g, W_g)$, i.e.,
\begin{equation}
    \tr(W_g^\dagger W) = \tr [\bar{Q}^\dagger (-R^\dagger W) + \bar{R}^\dagger W].
\end{equation}
We can then infer that the gradient is given by
\begin{equation}
    W_g = \bar{R} - R \bar{Q}.
\end{equation}

\paragraph{Retraction} After the search direction is determined, Riemannian optimization algorithm employs the concept of \emph{retraction} to travel along the search direction while staying within the manifold.
In our case, the retraction along the search direction $(-R^\dagger W, W)$ at point $(Q, R)$ is given by 
\begin{align}
    Q &\rightarrow Q - \alpha R^\dagger W - \frac{1}{2} \alpha^2 W^\dagger W, \\
    R &\rightarrow R + \alpha W. 
\end{align}
Here, $\alpha$ is the step length along the search direction, which is typically determined through a line search procedure. Note the additional $\alpha^2$ dependence which is needed to ensure that our retraction remains in the left-canocial form, and thus satisfies Eq.~\eqref{eq:left-canonical-condition}, beyond first order.

\paragraph{Vector transport} In the Riemannian optimization, to make use of information from the previous steps, we need to employ the concept of \emph{vector transport} to transport the tangent vectors (such as gradients) from previous steps to the current point. 
The vector transport should be compactible with the retraction scheme and the metric in the tangent space. 
In our case, we choose the vector transport to be the identity transformation $W \rightarrow W$. Indeed, that this is a valid choice is one of the main benefits of working with the unphysical inner product in Eq.~\eqref{eq:tangent-vector-dot}. Finding a vector transport that preserves the physical overlap would be much harder to construct.

\paragraph{Preconditioner} In the framework of L-BFGS, it is often beneficial to employ a preconditioner during the optimization. In particular, we can use this preconditioner to compensate for the fact that we have employed an unphysical inner product, rather than the natural inner product obtained from the physical overlap $\< \Phi(-R^\dagger W_1, W_1) | \Phi(-R^\dagger W_2, W_2)\>$, as was explored in Ref.~\cite{hauru-riemannian-2021} for the case of MPS and MERA. Suppose the physical overlap between tangent vectors have the form 
\begin{equation}
    \< \Phi(-R^\dagger W_1, W_1) | \Phi(-R^\dagger W_2, W_2)\> = \tr[W_1^\dagger W_2 \rho_W],
    \label{eq:physical-metric}
\end{equation}
where the matrix $\rho_W$ is a Hermitian, positive-(semi)definite matrix of size $\chi \times \chi$, which we will henceforth refer to as the physical metric. 
The preconditioner is then chosen to be a pseudo-inverse of $\rho_W$. 
When determining the search directions, we apply the following mapping to the gradient $(-R^\dagger W_g, W_g)$: 
\begin{equation}
    W_g \rightarrow W_g (\rho_W + \delta I)^{-1}.
\end{equation}
Here, $I$ is a $\chi\times \chi$ identity matrix, and $\delta$ is a small parameter which is chosen to be the norm of the original gradient.  

In the case of uniform circular cMPS, however, the straightforward application of Eq.~\eqref{eq:physical-metric} is difficult.
The computation of the physical overlap $\< \Phi(-R^\dagger W_1, W_1) | \Phi(-R^\dagger W_2, W_2)\>$ requires high computational cost and cannot be expressed in the format of Eq.~\eqref{eq:physical-metric}. Nevertheless, in practice, a reasonable approximation to the physical metric will suffice to accelerate the optimization process. 
We notice that, in the thermodynamic limit, the computation of the physical overlap is straightforward~\cite{draxler-particles-2013,hauru-riemannian-2021}
\begin{equation}
    \lim_{L \rightarrow \infty} \< \Phi(-R^\dagger W_1, W_1) | \Phi(-R^\dagger W_2, W_2)\> = \tr(W_1^\dagger W_2 \rho_R),
\end{equation} 
where $\rho_R$ is the dominant right eigenvector of the cMPS transfer matrix.
Since we are mainly interested in the cases where the system sizes are large, $\rho_R$ is a fairly reasonable approximation to $\rho_W$ is our computation.  

\begin{figure}[!htb]
    \centering
    \resizebox{\columnwidth}{!}{\includegraphics{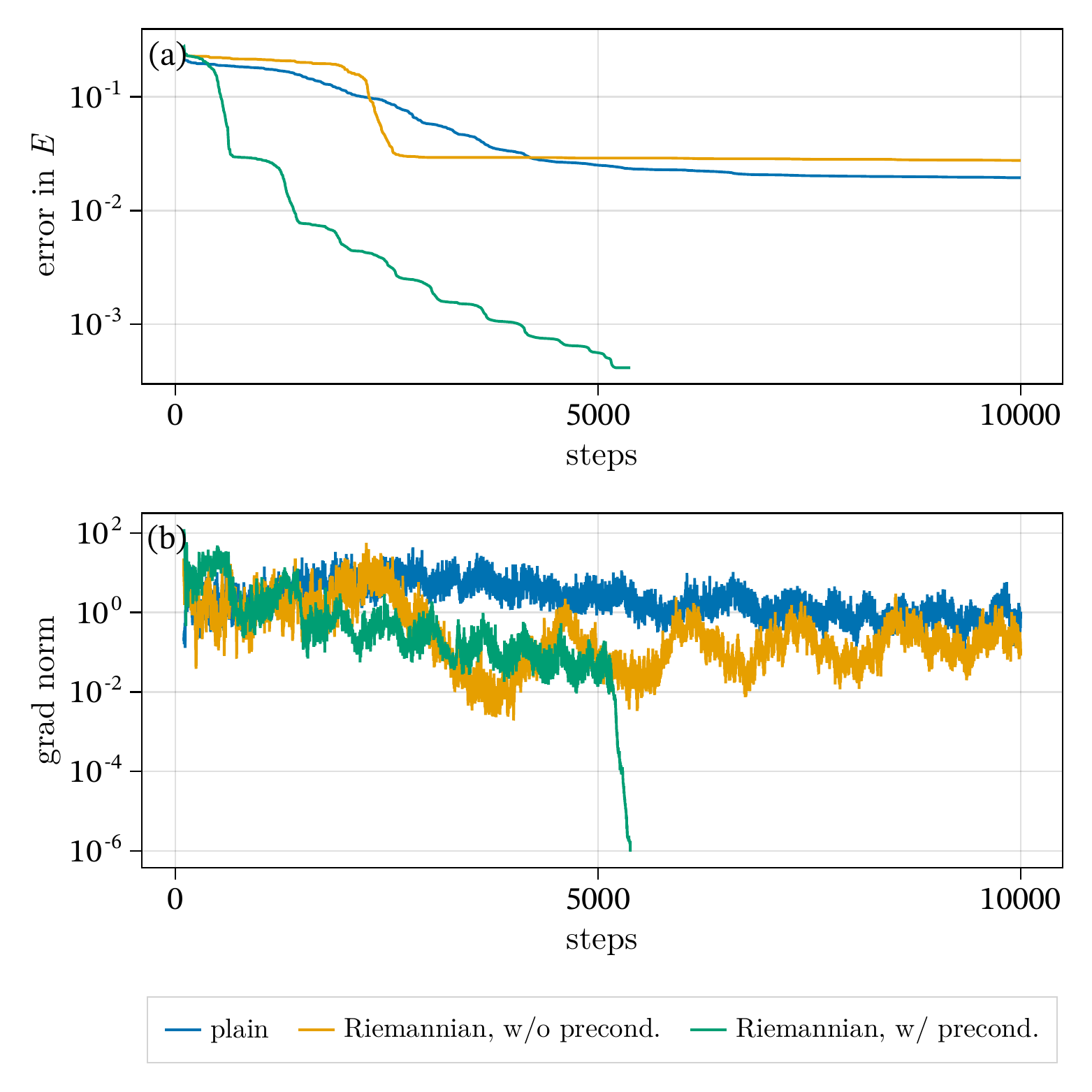}}
    \caption{Comparison of (a) the error in the energy and (b) the norm of the gradient versus the number of optimization steps for different optimization strategies. The first 100 optimization steps are omitted in the figure for the sake of clarity.}
    \label{fig:benchmark}
\end{figure}

\paragraph{Performance benchmark} Here, we present a performance benchmark of the Riemannian optimization method in an example of computing the ground state for the Lieb-Liniger Hamiltonian with $c = 1$, $\mu = 4$, and $L=32$.
The bond dimension of the cMPS is $\chi=12$. 

Starting from the same randomized initial state, we compare the performance of the following strategies: the plain L-BFGS optimization and the Riemannian generalization of L-BFGS algorithm (with and without the preconditioner).
As shown in Fig.~\ref{fig:benchmark}, the Riemannian optimization with the preconditioner clearly outperforms the other strategies, in which the preconditioner plays a crucial role. 

In practice, instead of directly optimizing the cMPS at the target bond dimension $\chi$, we usually start from a small bond dimension, and increase the bond dimension gradually.

\bibliography{liebliniger}

\begin{thebibliography}{64}%
\makeatletter
\providecommand \@ifxundefined [1]{%
 \@ifx{#1\undefined}
}%
\providecommand \@ifnum [1]{%
 \ifnum #1\expandafter \@firstoftwo
 \else \expandafter \@secondoftwo
 \fi
}%
\providecommand \@ifx [1]{%
 \ifx #1\expandafter \@firstoftwo
 \else \expandafter \@secondoftwo
 \fi
}%
\providecommand \natexlab [1]{#1}%
\providecommand \enquote  [1]{``#1''}%
\providecommand \bibnamefont  [1]{#1}%
\providecommand \bibfnamefont [1]{#1}%
\providecommand \citenamefont [1]{#1}%
\providecommand \href@noop [0]{\@secondoftwo}%
\providecommand \href [0]{\begingroup \@sanitize@url \@href}%
\providecommand \@href[1]{\@@startlink{#1}\@@href}%
\providecommand \@@href[1]{\endgroup#1\@@endlink}%
\providecommand \@sanitize@url [0]{\catcode `\\12\catcode `\$12\catcode `\&12\catcode `\#12\catcode `\^12\catcode `\_12\catcode `\%12\relax}%
\providecommand \@@startlink[1]{}%
\providecommand \@@endlink[0]{}%
\providecommand \url  [0]{\begingroup\@sanitize@url \@url }%
\providecommand \@url [1]{\endgroup\@href {#1}{\urlprefix }}%
\providecommand \urlprefix  [0]{URL }%
\providecommand \Eprint [0]{\href }%
\providecommand \doibase [0]{https://doi.org/}%
\providecommand \selectlanguage [0]{\@gobble}%
\providecommand \bibinfo  [0]{\@secondoftwo}%
\providecommand \bibfield  [0]{\@secondoftwo}%
\providecommand \translation [1]{[#1]}%
\providecommand \BibitemOpen [0]{}%
\providecommand \bibitemStop [0]{}%
\providecommand \bibitemNoStop [0]{.\EOS\space}%
\providecommand \EOS [0]{\spacefactor3000\relax}%
\providecommand \BibitemShut  [1]{\csname bibitem#1\endcsname}%
\let\auto@bib@innerbib\@empty
\bibitem [{\citenamefont {Henkel}(1999)}]{henkel-conformal-book-1999}%
  \BibitemOpen
  \bibfield  {author} {\bibinfo {author} {\bibfnamefont {M.}~\bibnamefont {Henkel}},\ }\href@noop {} {\emph {\bibinfo {title} {Conformal invariance and critical phenomena}}}\ (\bibinfo  {publisher} {Springer Science \& Business Media},\ \bibinfo {year} {1999})\BibitemShut {NoStop}%
\bibitem [{\citenamefont {Francesco}\ \emph {et~al.}(2012)\citenamefont {Francesco}, \citenamefont {Mathieu},\ and\ \citenamefont {S{\'e}n{\'e}chal}}]{francesco-conformal-book-2012}%
  \BibitemOpen
  \bibfield  {author} {\bibinfo {author} {\bibfnamefont {P.}~\bibnamefont {Francesco}}, \bibinfo {author} {\bibfnamefont {P.}~\bibnamefont {Mathieu}},\ and\ \bibinfo {author} {\bibfnamefont {D.}~\bibnamefont {S{\'e}n{\'e}chal}},\ }\href@noop {} {\emph {\bibinfo {title} {Conformal field theory}}}\ (\bibinfo  {publisher} {Springer Science \& Business Media},\ \bibinfo {year} {2012})\BibitemShut {NoStop}%
\bibitem [{\citenamefont {Cardy}(1984)}]{cardy-conformal-1984}%
  \BibitemOpen
  \bibfield  {author} {\bibinfo {author} {\bibfnamefont {J.~L.}\ \bibnamefont {Cardy}},\ }\href {https://doi.org/10.1088/0305-4470/17/7/003} {\bibfield  {journal} {\bibinfo  {journal} {J. Phys. A: Math. Gen.}\ }\textbf {\bibinfo {volume} {17}},\ \bibinfo {pages} {L385} (\bibinfo {year} {1984})}\BibitemShut {NoStop}%
\bibitem [{\citenamefont {Bl\"ote}\ \emph {et~al.}(1986)\citenamefont {Bl\"ote}, \citenamefont {Cardy},\ and\ \citenamefont {Nightingale}}]{blote-conformal-1986}%
  \BibitemOpen
  \bibfield  {author} {\bibinfo {author} {\bibfnamefont {H.~W.~J.}\ \bibnamefont {Bl\"ote}}, \bibinfo {author} {\bibfnamefont {J.~L.}\ \bibnamefont {Cardy}},\ and\ \bibinfo {author} {\bibfnamefont {M.~P.}\ \bibnamefont {Nightingale}},\ }\href {https://doi.org/10.1103/PhysRevLett.56.742} {\bibfield  {journal} {\bibinfo  {journal} {Phys. Rev. Lett.}\ }\textbf {\bibinfo {volume} {56}},\ \bibinfo {pages} {742} (\bibinfo {year} {1986})}\BibitemShut {NoStop}%
\bibitem [{\citenamefont {Cardy}(1986{\natexlab{a}})}]{cardy-logarithmic-1986}%
  \BibitemOpen
  \bibfield  {author} {\bibinfo {author} {\bibfnamefont {J.~L.}\ \bibnamefont {Cardy}},\ }\href {https://iopscience.iop.org/article/10.1088/0305-4470/19/17/008} {\bibfield  {journal} {\bibinfo  {journal} {J. Phys. A: Math. Gen.}\ }\textbf {\bibinfo {volume} {19}},\ \bibinfo {pages} {L1093} (\bibinfo {year} {1986}{\natexlab{a}})}\BibitemShut {NoStop}%
\bibitem [{\citenamefont {Cardy}(1986{\natexlab{b}})}]{cardy-operator-1986}%
  \BibitemOpen
  \bibfield  {author} {\bibinfo {author} {\bibfnamefont {J.~L.}\ \bibnamefont {Cardy}},\ }\href {https://www.sciencedirect.com/science/article/abs/pii/0550321386905523?via%3Dihub} {\bibfield  {journal} {\bibinfo  {journal} {Nucl. Phys. B}\ }\textbf {\bibinfo {volume} {270}},\ \bibinfo {pages} {186} (\bibinfo {year} {1986}{\natexlab{b}})}\BibitemShut {NoStop}%
\bibitem [{\citenamefont {Affleck}(1986)}]{affleck-universal-1986}%
  \BibitemOpen
  \bibfield  {author} {\bibinfo {author} {\bibfnamefont {I.}~\bibnamefont {Affleck}},\ }\href {https://doi.org/10.1103/PhysRevLett.56.746} {\bibfield  {journal} {\bibinfo  {journal} {Phys. Rev. Lett.}\ }\textbf {\bibinfo {volume} {56}},\ \bibinfo {pages} {746} (\bibinfo {year} {1986})}\BibitemShut {NoStop}%
\bibitem [{\citenamefont {Koo}\ and\ \citenamefont {Saleur}(1994)}]{koo-representations-1994}%
  \BibitemOpen
  \bibfield  {author} {\bibinfo {author} {\bibfnamefont {W.}~\bibnamefont {Koo}}\ and\ \bibinfo {author} {\bibfnamefont {H.}~\bibnamefont {Saleur}},\ }\href {https://www.sciencedirect.com/science/article/abs/pii/0550321394900183?via%3Dihub} {\bibfield  {journal} {\bibinfo  {journal} {Nucl. Phys. B}\ }\textbf {\bibinfo {volume} {426}},\ \bibinfo {pages} {459} (\bibinfo {year} {1994})}\BibitemShut {NoStop}%
\bibitem [{\citenamefont {Milsted}\ and\ \citenamefont {Vidal}(2017)}]{milsted-extraction-2017}%
  \BibitemOpen
  \bibfield  {author} {\bibinfo {author} {\bibfnamefont {A.}~\bibnamefont {Milsted}}\ and\ \bibinfo {author} {\bibfnamefont {G.}~\bibnamefont {Vidal}},\ }\href {https://doi.org/10.1103/PhysRevB.96.245105} {\bibfield  {journal} {\bibinfo  {journal} {Phys. Rev. B}\ }\textbf {\bibinfo {volume} {96}},\ \bibinfo {pages} {245105} (\bibinfo {year} {2017})}\BibitemShut {NoStop}%
\bibitem [{\citenamefont {Pirvu}\ \emph {et~al.}(2011)\citenamefont {Pirvu}, \citenamefont {Verstraete},\ and\ \citenamefont {Vidal}}]{pirvu-exploiting-2011}%
  \BibitemOpen
  \bibfield  {author} {\bibinfo {author} {\bibfnamefont {B.}~\bibnamefont {Pirvu}}, \bibinfo {author} {\bibfnamefont {F.}~\bibnamefont {Verstraete}},\ and\ \bibinfo {author} {\bibfnamefont {G.}~\bibnamefont {Vidal}},\ }\href {https://doi.org/10.1103/PhysRevB.83.125104} {\bibfield  {journal} {\bibinfo  {journal} {Phys. Rev. B}\ }\textbf {\bibinfo {volume} {83}},\ \bibinfo {pages} {125104} (\bibinfo {year} {2011})}\BibitemShut {NoStop}%
\bibitem [{\citenamefont {Haegeman}\ \emph {et~al.}(2012)\citenamefont {Haegeman}, \citenamefont {Pirvu}, \citenamefont {Weir}, \citenamefont {Cirac}, \citenamefont {Osborne}, \citenamefont {Verschelde},\ and\ \citenamefont {Verstraete}}]{haegeman-variational-2012}%
  \BibitemOpen
  \bibfield  {author} {\bibinfo {author} {\bibfnamefont {J.}~\bibnamefont {Haegeman}}, \bibinfo {author} {\bibfnamefont {B.}~\bibnamefont {Pirvu}}, \bibinfo {author} {\bibfnamefont {D.~J.}\ \bibnamefont {Weir}}, \bibinfo {author} {\bibfnamefont {J.~I.}\ \bibnamefont {Cirac}}, \bibinfo {author} {\bibfnamefont {T.~J.}\ \bibnamefont {Osborne}}, \bibinfo {author} {\bibfnamefont {H.}~\bibnamefont {Verschelde}},\ and\ \bibinfo {author} {\bibfnamefont {F.}~\bibnamefont {Verstraete}},\ }\href {https://doi.org/10.1103/PhysRevB.85.100408} {\bibfield  {journal} {\bibinfo  {journal} {Phys. Rev. B}\ }\textbf {\bibinfo {volume} {85}},\ \bibinfo {pages} {100408} (\bibinfo {year} {2012})}\BibitemShut {NoStop}%
\bibitem [{\citenamefont {Pirvu}\ \emph {et~al.}(2012)\citenamefont {Pirvu}, \citenamefont {Haegeman},\ and\ \citenamefont {Verstraete}}]{pirvu-matrix-2012-b}%
  \BibitemOpen
  \bibfield  {author} {\bibinfo {author} {\bibfnamefont {B.}~\bibnamefont {Pirvu}}, \bibinfo {author} {\bibfnamefont {J.}~\bibnamefont {Haegeman}},\ and\ \bibinfo {author} {\bibfnamefont {F.}~\bibnamefont {Verstraete}},\ }\href {https://doi.org/10.1103/PhysRevB.85.035130} {\bibfield  {journal} {\bibinfo  {journal} {Phys. Rev. B}\ }\textbf {\bibinfo {volume} {85}},\ \bibinfo {pages} {035130} (\bibinfo {year} {2012})}\BibitemShut {NoStop}%
\bibitem [{\citenamefont {Zou}\ \emph {et~al.}(2018)\citenamefont {Zou}, \citenamefont {Milsted},\ and\ \citenamefont {Vidal}}]{zou-conformal-2018}%
  \BibitemOpen
  \bibfield  {author} {\bibinfo {author} {\bibfnamefont {Y.}~\bibnamefont {Zou}}, \bibinfo {author} {\bibfnamefont {A.}~\bibnamefont {Milsted}},\ and\ \bibinfo {author} {\bibfnamefont {G.}~\bibnamefont {Vidal}},\ }\href {https://doi.org/10.1103/PhysRevLett.121.230402} {\bibfield  {journal} {\bibinfo  {journal} {Phys. Rev. Lett.}\ }\textbf {\bibinfo {volume} {121}},\ \bibinfo {pages} {230402} (\bibinfo {year} {2018})}\BibitemShut {NoStop}%
\bibitem [{\citenamefont {Zou}\ \emph {et~al.}(2020)\citenamefont {Zou}, \citenamefont {Milsted},\ and\ \citenamefont {Vidal}}]{zou-conformal-2020}%
  \BibitemOpen
  \bibfield  {author} {\bibinfo {author} {\bibfnamefont {Y.}~\bibnamefont {Zou}}, \bibinfo {author} {\bibfnamefont {A.}~\bibnamefont {Milsted}},\ and\ \bibinfo {author} {\bibfnamefont {G.}~\bibnamefont {Vidal}},\ }\href {https://doi.org/10.1103/PhysRevLett.124.040604} {\bibfield  {journal} {\bibinfo  {journal} {Phys. Rev. Lett.}\ }\textbf {\bibinfo {volume} {124}},\ \bibinfo {pages} {040604} (\bibinfo {year} {2020})}\BibitemShut {NoStop}%
\bibitem [{\citenamefont {Zou}\ and\ \citenamefont {Vidal}(2020)}]{zou-emergence-2020}%
  \BibitemOpen
  \bibfield  {author} {\bibinfo {author} {\bibfnamefont {Y.}~\bibnamefont {Zou}}\ and\ \bibinfo {author} {\bibfnamefont {G.}~\bibnamefont {Vidal}},\ }\href {https://doi.org/10.1103/PhysRevB.101.045132} {\bibfield  {journal} {\bibinfo  {journal} {Phys. Rev. B}\ }\textbf {\bibinfo {volume} {101}},\ \bibinfo {pages} {045132} (\bibinfo {year} {2020})}\BibitemShut {NoStop}%
\bibitem [{\citenamefont {Blumenhagen}\ and\ \citenamefont {Plauschinn}(2009)}]{blumenhagen-introduction-2009}%
  \BibitemOpen
  \bibfield  {author} {\bibinfo {author} {\bibfnamefont {R.}~\bibnamefont {Blumenhagen}}\ and\ \bibinfo {author} {\bibfnamefont {E.}~\bibnamefont {Plauschinn}},\ }\href@noop {} {\emph {\bibinfo {title} {Introduction to conformal field theory: with applications to string theory}}},\ Vol.\ \bibinfo {volume} {779}\ (\bibinfo  {publisher} {Springer, Berlin, Heidelberg},\ \bibinfo {year} {2009})\BibitemShut {NoStop}%
\bibitem [{\citenamefont {Mussardo}(2020)}]{mussardo-statistical-2020}%
  \BibitemOpen
  \bibfield  {author} {\bibinfo {author} {\bibfnamefont {G.}~\bibnamefont {Mussardo}},\ }\href {https://global.oup.com/academic/product/statistical-field-theory-9780198788102?cc=cn&lang=en&} {\emph {\bibinfo {title} {Statistical {Field} {Theory}: {An} {Introduction} to {Exactly} {Solved} {Models} in {Statistical} {Physics}}}},\ \bibinfo {edition} {second edition}\ ed.,\ Oxford {Graduate} {Texts}\ (\bibinfo  {publisher} {Oxford University Press},\ \bibinfo {address} {Oxford, New York},\ \bibinfo {year} {2020})\BibitemShut {NoStop}%
\bibitem [{\citenamefont {Wang}\ \emph {et~al.}(2022)\citenamefont {Wang}, \citenamefont {Zou},\ and\ \citenamefont {Vidal}}]{wang-emergence-2022}%
  \BibitemOpen
  \bibfield  {author} {\bibinfo {author} {\bibfnamefont {R.}~\bibnamefont {Wang}}, \bibinfo {author} {\bibfnamefont {Y.}~\bibnamefont {Zou}},\ and\ \bibinfo {author} {\bibfnamefont {G.}~\bibnamefont {Vidal}},\ }\href {https://doi.org/10.1103/PhysRevB.106.115111} {\bibfield  {journal} {\bibinfo  {journal} {Phys. Rev. B}\ }\textbf {\bibinfo {volume} {106}},\ \bibinfo {pages} {115111} (\bibinfo {year} {2022})}\BibitemShut {NoStop}%
\bibitem [{\citenamefont {Yang}\ \emph {et~al.}(2022)\citenamefont {Yang}, \citenamefont {Vanhecke},\ and\ \citenamefont {Schuch}}]{yang-detecting-2022}%
  \BibitemOpen
  \bibfield  {author} {\bibinfo {author} {\bibfnamefont {M.}~\bibnamefont {Yang}}, \bibinfo {author} {\bibfnamefont {B.}~\bibnamefont {Vanhecke}},\ and\ \bibinfo {author} {\bibfnamefont {N.}~\bibnamefont {Schuch}},\ }\href {https://arxiv.org/abs/2210.17539} {\bibfield  {journal} {\bibinfo  {journal} {arXiv:2210.17539}\ } (\bibinfo {year} {2022})}\BibitemShut {NoStop}%
\bibitem [{\citenamefont {Haldane}(1981{\natexlab{a}})}]{haldane-effective-1981}%
  \BibitemOpen
  \bibfield  {author} {\bibinfo {author} {\bibfnamefont {F.~D.~M.}\ \bibnamefont {Haldane}},\ }\href {https://doi.org/10.1103/PhysRevLett.47.1840} {\bibfield  {journal} {\bibinfo  {journal} {Phys. Rev. Lett.}\ }\textbf {\bibinfo {volume} {47}},\ \bibinfo {pages} {1840} (\bibinfo {year} {1981}{\natexlab{a}})}\BibitemShut {NoStop}%
\bibitem [{\citenamefont {Cazalilla}(2004)}]{cazalilla-bosonizing-2004}%
  \BibitemOpen
  \bibfield  {author} {\bibinfo {author} {\bibfnamefont {M.}~\bibnamefont {Cazalilla}},\ }\href {https://iopscience.iop.org/article/10.1088/0953-4075/37/7/051} {\bibfield  {journal} {\bibinfo  {journal} {J. Phys. B: At. Mol. Opt. Phys.}\ }\textbf {\bibinfo {volume} {37}},\ \bibinfo {pages} {S1} (\bibinfo {year} {2004})}\BibitemShut {NoStop}%
\bibitem [{\citenamefont {Lieb}\ and\ \citenamefont {Liniger}(1963)}]{lieb-exact-1963-a}%
  \BibitemOpen
  \bibfield  {author} {\bibinfo {author} {\bibfnamefont {E.~H.}\ \bibnamefont {Lieb}}\ and\ \bibinfo {author} {\bibfnamefont {W.}~\bibnamefont {Liniger}},\ }\href {https://doi.org/10.1103/PhysRev.130.1605} {\bibfield  {journal} {\bibinfo  {journal} {Phys. Rev.}\ }\textbf {\bibinfo {volume} {130}},\ \bibinfo {pages} {1605} (\bibinfo {year} {1963})}\BibitemShut {NoStop}%
\bibitem [{\citenamefont {Lieb}(1963)}]{lieb-exact-1963-b}%
  \BibitemOpen
  \bibfield  {author} {\bibinfo {author} {\bibfnamefont {E.~H.}\ \bibnamefont {Lieb}},\ }\href {https://doi.org/10.1103/PhysRev.130.1616} {\bibfield  {journal} {\bibinfo  {journal} {Phys. Rev.}\ }\textbf {\bibinfo {volume} {130}},\ \bibinfo {pages} {1616} (\bibinfo {year} {1963})}\BibitemShut {NoStop}%
\bibitem [{\citenamefont {Korepin}\ \emph {et~al.}(1997)\citenamefont {Korepin}, \citenamefont {Bogoliubov},\ and\ \citenamefont {Izergin}}]{korepin-quantum-book-1997}%
  \BibitemOpen
  \bibfield  {author} {\bibinfo {author} {\bibfnamefont {V.~E.}\ \bibnamefont {Korepin}}, \bibinfo {author} {\bibfnamefont {N.~M.}\ \bibnamefont {Bogoliubov}},\ and\ \bibinfo {author} {\bibfnamefont {A.~G.}\ \bibnamefont {Izergin}},\ }\href@noop {} {\emph {\bibinfo {title} {Quantum inverse scattering method and correlation functions}}}\ (\bibinfo  {publisher} {Cambridge university press},\ \bibinfo {year} {1997})\BibitemShut {NoStop}%
\bibitem [{\citenamefont {Gaudin}(2014)}]{gaudin-bethe-book-2014}%
  \BibitemOpen
  \bibfield  {author} {\bibinfo {author} {\bibfnamefont {M.}~\bibnamefont {Gaudin}},\ }\href@noop {} {\emph {\bibinfo {title} {The Bethe Wavefunction}}}\ (\bibinfo  {publisher} {Cambridge University Press},\ \bibinfo {year} {2014})\BibitemShut {NoStop}%
\bibitem [{\citenamefont {Yang}\ and\ \citenamefont {Yang}(1969)}]{yang-thermodynamics-1969}%
  \BibitemOpen
  \bibfield  {author} {\bibinfo {author} {\bibfnamefont {C.-N.}\ \bibnamefont {Yang}}\ and\ \bibinfo {author} {\bibfnamefont {C.~P.}\ \bibnamefont {Yang}},\ }\href {https://aip.scitation.org/doi/10.1063/1.1664947} {\bibfield  {journal} {\bibinfo  {journal} {J. Math. Phys.}\ }\textbf {\bibinfo {volume} {10}},\ \bibinfo {pages} {1115} (\bibinfo {year} {1969})}\BibitemShut {NoStop}%
\bibitem [{\citenamefont {Jiang}\ \emph {et~al.}(2015)\citenamefont {Jiang}, \citenamefont {Chen},\ and\ \citenamefont {Guan}}]{jiang-understanding-2015}%
  \BibitemOpen
  \bibfield  {author} {\bibinfo {author} {\bibfnamefont {Y.-Z.}\ \bibnamefont {Jiang}}, \bibinfo {author} {\bibfnamefont {Y.-Y.}\ \bibnamefont {Chen}},\ and\ \bibinfo {author} {\bibfnamefont {X.-W.}\ \bibnamefont {Guan}},\ }\href {https://doi.org/10.1088/1674-1056/24/5/050311} {\bibfield  {journal} {\bibinfo  {journal} {Chin. Phys. B}\ }\textbf {\bibinfo {volume} {24}},\ \bibinfo {pages} {050311} (\bibinfo {year} {2015})}\BibitemShut {NoStop}%
\bibitem [{\citenamefont {Verstraete}\ and\ \citenamefont {Cirac}(2010)}]{verstraete-continuous-2010}%
  \BibitemOpen
  \bibfield  {author} {\bibinfo {author} {\bibfnamefont {F.}~\bibnamefont {Verstraete}}\ and\ \bibinfo {author} {\bibfnamefont {J.~I.}\ \bibnamefont {Cirac}},\ }\href {https://doi.org/10.1103/PhysRevLett.104.190405} {\bibfield  {journal} {\bibinfo  {journal} {Phys. Rev. Lett.}\ }\textbf {\bibinfo {volume} {104}},\ \bibinfo {pages} {190405} (\bibinfo {year} {2010})}\BibitemShut {NoStop}%
\bibitem [{\citenamefont {Haegeman}\ \emph {et~al.}(2013)\citenamefont {Haegeman}, \citenamefont {Cirac}, \citenamefont {Osborne},\ and\ \citenamefont {Verstraete}}]{haegeman-calculus-2013}%
  \BibitemOpen
  \bibfield  {author} {\bibinfo {author} {\bibfnamefont {J.}~\bibnamefont {Haegeman}}, \bibinfo {author} {\bibfnamefont {J.~I.}\ \bibnamefont {Cirac}}, \bibinfo {author} {\bibfnamefont {T.~J.}\ \bibnamefont {Osborne}},\ and\ \bibinfo {author} {\bibfnamefont {F.}~\bibnamefont {Verstraete}},\ }\href {https://journals.aps.org/prb/abstract/10.1103/PhysRevB.88.085118} {\bibfield  {journal} {\bibinfo  {journal} {Phys. Rev. B}\ }\textbf {\bibinfo {volume} {88}},\ \bibinfo {pages} {085118} (\bibinfo {year} {2013})}\BibitemShut {NoStop}%
\bibitem [{\citenamefont {Haldane}(1981{\natexlab{b}})}]{haldane-luttinger-1981}%
  \BibitemOpen
  \bibfield  {author} {\bibinfo {author} {\bibfnamefont {F.}~\bibnamefont {Haldane}},\ }\href {https://iopscience.iop.org/article/10.1088/0022-3719/14/19/010} {\bibfield  {journal} {\bibinfo  {journal} {J. Phys. C: Solid State Phys.}\ }\textbf {\bibinfo {volume} {14}},\ \bibinfo {pages} {2585} (\bibinfo {year} {1981}{\natexlab{b}})}\BibitemShut {NoStop}%
\bibitem [{\citenamefont {Ludwig}(1995)}]{ludwig-methods-1995}%
  \BibitemOpen
  \bibfield  {author} {\bibinfo {author} {\bibfnamefont {A.~W.~W.}\ \bibnamefont {Ludwig}},\ }in\ \href {https://www.worldscientific.com/worldscibooks/10.1142/2634#t=aboutBook} {\emph {\bibinfo {booktitle} {Low-dimensional quantum field theories for condensed matter physicists}}}\ (\bibinfo  {publisher} {World Scientific},\ \bibinfo {address} {Trieste, Italy},\ \bibinfo {year} {1995})\ pp.\ \bibinfo {pages} {389--455}\BibitemShut {NoStop}%
\bibitem [{\citenamefont {von Delft}\ and\ \citenamefont {Schoeller}(1998)}]{vondelft-bosonization-1998}%
  \BibitemOpen
  \bibfield  {author} {\bibinfo {author} {\bibfnamefont {J.}~\bibnamefont {von Delft}}\ and\ \bibinfo {author} {\bibfnamefont {H.}~\bibnamefont {Schoeller}},\ }\href {https://onlinelibrary.wiley.com/doi/10.1002/andp.19985100401} {\bibfield  {journal} {\bibinfo  {journal} {Ann. Phys.}\ }\textbf {\bibinfo {volume} {7}},\ \bibinfo {pages} {225} (\bibinfo {year} {1998})}\BibitemShut {NoStop}%
\bibitem [{\citenamefont {Slavnov}(1990)}]{slavnov-nonequal-1990}%
  \BibitemOpen
  \bibfield  {author} {\bibinfo {author} {\bibfnamefont {N.~A.}\ \bibnamefont {Slavnov}},\ }\href {https://link.springer.com/article/10.1007/BF01029221} {\bibfield  {journal} {\bibinfo  {journal} {Theor. Math. Phys.}\ }\textbf {\bibinfo {volume} {82}},\ \bibinfo {pages} {273} (\bibinfo {year} {1990})}\BibitemShut {NoStop}%
\bibitem [{\citenamefont {De~Nardis}\ and\ \citenamefont {Panfil}(2015)}]{de-nardis-density-2015}%
  \BibitemOpen
  \bibfield  {author} {\bibinfo {author} {\bibfnamefont {J.}~\bibnamefont {De~Nardis}}\ and\ \bibinfo {author} {\bibfnamefont {M.}~\bibnamefont {Panfil}},\ }\href {https://iopscience.iop.org/article/10.1088/1742-5468/2015/02/P02019} {\bibfield  {journal} {\bibinfo  {journal} {J. Stat. Mech.}\ }\textbf {\bibinfo {volume} {2015}},\ \bibinfo {pages} {P02019} (\bibinfo {year} {2015})}\BibitemShut {NoStop}%
\bibitem [{\citenamefont {Bouchoule}\ \emph {et~al.}(2022)\citenamefont {Bouchoule}, \citenamefont {Dubail}, \citenamefont {Dubois},\ and\ \citenamefont {Gangardt}}]{bouchoule-relaxation-2022}%
  \BibitemOpen
  \bibfield  {author} {\bibinfo {author} {\bibfnamefont {I.}~\bibnamefont {Bouchoule}}, \bibinfo {author} {\bibfnamefont {J.}~\bibnamefont {Dubail}}, \bibinfo {author} {\bibfnamefont {L.}~\bibnamefont {Dubois}},\ and\ \bibinfo {author} {\bibfnamefont {D.~M.}\ \bibnamefont {Gangardt}},\ }\href {https://arxiv.org/abs/2206.00112} {\bibfield  {journal} {\bibinfo  {journal} {arXiv:2206.00112}\ } (\bibinfo {year} {2022})}\BibitemShut {NoStop}%
\bibitem [{\citenamefont {Haegeman}\ \emph {et~al.}(2010)\citenamefont {Haegeman}, \citenamefont {Cirac}, \citenamefont {Osborne}, \citenamefont {Verschelde},\ and\ \citenamefont {Verstraete}}]{haegeman-applying-2010}%
  \BibitemOpen
  \bibfield  {author} {\bibinfo {author} {\bibfnamefont {J.}~\bibnamefont {Haegeman}}, \bibinfo {author} {\bibfnamefont {J.~I.}\ \bibnamefont {Cirac}}, \bibinfo {author} {\bibfnamefont {T.~J.}\ \bibnamefont {Osborne}}, \bibinfo {author} {\bibfnamefont {H.}~\bibnamefont {Verschelde}},\ and\ \bibinfo {author} {\bibfnamefont {F.}~\bibnamefont {Verstraete}},\ }\href {https://doi.org/10.1103/PhysRevLett.105.251601} {\bibfield  {journal} {\bibinfo  {journal} {Phys. Rev. Lett.}\ }\textbf {\bibinfo {volume} {105}},\ \bibinfo {pages} {251601} (\bibinfo {year} {2010})}\BibitemShut {NoStop}%
\bibitem [{\citenamefont {Rinc\'on}\ \emph {et~al.}(2015)\citenamefont {Rinc\'on}, \citenamefont {Ganahl},\ and\ \citenamefont {Vidal}}]{rincon-liebliniger-2015}%
  \BibitemOpen
  \bibfield  {author} {\bibinfo {author} {\bibfnamefont {J.}~\bibnamefont {Rinc\'on}}, \bibinfo {author} {\bibfnamefont {M.}~\bibnamefont {Ganahl}},\ and\ \bibinfo {author} {\bibfnamefont {G.}~\bibnamefont {Vidal}},\ }\href {https://doi.org/10.1103/PhysRevB.92.115107} {\bibfield  {journal} {\bibinfo  {journal} {Phys. Rev. B}\ }\textbf {\bibinfo {volume} {92}},\ \bibinfo {pages} {115107} (\bibinfo {year} {2015})}\BibitemShut {NoStop}%
\bibitem [{\citenamefont {Ganahl}\ \emph {et~al.}(2017)\citenamefont {Ganahl}, \citenamefont {Rinc\'on},\ and\ \citenamefont {Vidal}}]{ganahl-continuous-2017-a}%
  \BibitemOpen
  \bibfield  {author} {\bibinfo {author} {\bibfnamefont {M.}~\bibnamefont {Ganahl}}, \bibinfo {author} {\bibfnamefont {J.}~\bibnamefont {Rinc\'on}},\ and\ \bibinfo {author} {\bibfnamefont {G.}~\bibnamefont {Vidal}},\ }\href {https://doi.org/10.1103/PhysRevLett.118.220402} {\bibfield  {journal} {\bibinfo  {journal} {Phys. Rev. Lett.}\ }\textbf {\bibinfo {volume} {118}},\ \bibinfo {pages} {220402} (\bibinfo {year} {2017})}\BibitemShut {NoStop}%
\bibitem [{\citenamefont {Tuybens}\ \emph {et~al.}(2022)\citenamefont {Tuybens}, \citenamefont {De~Nardis}, \citenamefont {Haegeman},\ and\ \citenamefont {Verstraete}}]{tuybens-variational-2022}%
  \BibitemOpen
  \bibfield  {author} {\bibinfo {author} {\bibfnamefont {B.}~\bibnamefont {Tuybens}}, \bibinfo {author} {\bibfnamefont {J.}~\bibnamefont {De~Nardis}}, \bibinfo {author} {\bibfnamefont {J.}~\bibnamefont {Haegeman}},\ and\ \bibinfo {author} {\bibfnamefont {F.}~\bibnamefont {Verstraete}},\ }\href {https://doi.org/10.1103/PhysRevLett.128.020501} {\bibfield  {journal} {\bibinfo  {journal} {Phys. Rev. Lett.}\ }\textbf {\bibinfo {volume} {128}},\ \bibinfo {pages} {020501} (\bibinfo {year} {2022})}\BibitemShut {NoStop}%
\bibitem [{\citenamefont {Lukin}\ and\ \citenamefont {Sotnikov}(2022)}]{lukin-continuous-2022}%
  \BibitemOpen
  \bibfield  {author} {\bibinfo {author} {\bibfnamefont {I.~V.}\ \bibnamefont {Lukin}}\ and\ \bibinfo {author} {\bibfnamefont {A.~G.}\ \bibnamefont {Sotnikov}},\ }\href {https://doi.org/10.1103/PhysRevB.106.144206} {\bibfield  {journal} {\bibinfo  {journal} {Phys. Rev. B}\ }\textbf {\bibinfo {volume} {106}},\ \bibinfo {pages} {144206} (\bibinfo {year} {2022})}\BibitemShut {NoStop}%
\bibitem [{\citenamefont {Draxler}\ \emph {et~al.}(2013)\citenamefont {Draxler}, \citenamefont {Haegeman}, \citenamefont {Osborne}, \citenamefont {Stojevic}, \citenamefont {Vanderstraeten},\ and\ \citenamefont {Verstraete}}]{draxler-particles-2013}%
  \BibitemOpen
  \bibfield  {author} {\bibinfo {author} {\bibfnamefont {D.}~\bibnamefont {Draxler}}, \bibinfo {author} {\bibfnamefont {J.}~\bibnamefont {Haegeman}}, \bibinfo {author} {\bibfnamefont {T.~J.}\ \bibnamefont {Osborne}}, \bibinfo {author} {\bibfnamefont {V.}~\bibnamefont {Stojevic}}, \bibinfo {author} {\bibfnamefont {L.}~\bibnamefont {Vanderstraeten}},\ and\ \bibinfo {author} {\bibfnamefont {F.}~\bibnamefont {Verstraete}},\ }\href {https://doi.org/10.1103/PhysRevLett.111.020402} {\bibfield  {journal} {\bibinfo  {journal} {Phys. Rev. Lett.}\ }\textbf {\bibinfo {volume} {111}},\ \bibinfo {pages} {020402} (\bibinfo {year} {2013})}\BibitemShut {NoStop}%
\bibitem [{\citenamefont {Haegeman}\ \emph {et~al.}(2017)\citenamefont {Haegeman}, \citenamefont {Draxler}, \citenamefont {Stojevic}, \citenamefont {Cirac}, \citenamefont {Osborne},\ and\ \citenamefont {Verstraete}}]{haegeman-quantum-2017}%
  \BibitemOpen
  \bibfield  {author} {\bibinfo {author} {\bibfnamefont {J.}~\bibnamefont {Haegeman}}, \bibinfo {author} {\bibfnamefont {D.}~\bibnamefont {Draxler}}, \bibinfo {author} {\bibfnamefont {V.}~\bibnamefont {Stojevic}}, \bibinfo {author} {\bibfnamefont {I.}~\bibnamefont {Cirac}}, \bibinfo {author} {\bibfnamefont {T.}~\bibnamefont {Osborne}},\ and\ \bibinfo {author} {\bibfnamefont {F.}~\bibnamefont {Verstraete}},\ }\href {https://www.scipost.org/SciPostPhys.3.1.006?acad_field_slug=physics} {\bibfield  {journal} {\bibinfo  {journal} {SciPost Phys.}\ }\textbf {\bibinfo {volume} {3}},\ \bibinfo {pages} {006} (\bibinfo {year} {2017})}\BibitemShut {NoStop}%
\bibitem [{\citenamefont {Draxler}\ \emph {et~al.}(2017)\citenamefont {Draxler}, \citenamefont {Haegeman}, \citenamefont {Verstraete},\ and\ \citenamefont {Rizzi}}]{draxler-continuous-2017}%
  \BibitemOpen
  \bibfield  {author} {\bibinfo {author} {\bibfnamefont {D.}~\bibnamefont {Draxler}}, \bibinfo {author} {\bibfnamefont {J.}~\bibnamefont {Haegeman}}, \bibinfo {author} {\bibfnamefont {F.}~\bibnamefont {Verstraete}},\ and\ \bibinfo {author} {\bibfnamefont {M.}~\bibnamefont {Rizzi}},\ }\href {https://doi.org/10.1103/PhysRevB.95.045145} {\bibfield  {journal} {\bibinfo  {journal} {Phys. Rev. B}\ }\textbf {\bibinfo {volume} {95}},\ \bibinfo {pages} {045145} (\bibinfo {year} {2017})}\BibitemShut {NoStop}%
\bibitem [{\citenamefont {Osborne}\ \emph {et~al.}(2010)\citenamefont {Osborne}, \citenamefont {Eisert},\ and\ \citenamefont {Verstraete}}]{osborne-holographic-2010}%
  \BibitemOpen
  \bibfield  {author} {\bibinfo {author} {\bibfnamefont {T.~J.}\ \bibnamefont {Osborne}}, \bibinfo {author} {\bibfnamefont {J.}~\bibnamefont {Eisert}},\ and\ \bibinfo {author} {\bibfnamefont {F.}~\bibnamefont {Verstraete}},\ }\href {https://doi.org/10.1103/PhysRevLett.105.260401} {\bibfield  {journal} {\bibinfo  {journal} {Phys. Rev. Lett.}\ }\textbf {\bibinfo {volume} {105}},\ \bibinfo {pages} {260401} (\bibinfo {year} {2010})}\BibitemShut {NoStop}%
\bibitem [{\citenamefont {Kiukas}\ \emph {et~al.}(2015)\citenamefont {Kiukas}, \citenamefont {Gu\c{t}\u{a}}, \citenamefont {Lesanovsky},\ and\ \citenamefont {Garrahan}}]{kiukas-equivalence-2015}%
  \BibitemOpen
  \bibfield  {author} {\bibinfo {author} {\bibfnamefont {J.}~\bibnamefont {Kiukas}}, \bibinfo {author} {\bibfnamefont {M.}~\bibnamefont {Gu\c{t}\u{a}}}, \bibinfo {author} {\bibfnamefont {I.}~\bibnamefont {Lesanovsky}},\ and\ \bibinfo {author} {\bibfnamefont {J.~P.}\ \bibnamefont {Garrahan}},\ }\href {https://doi.org/10.1103/PhysRevE.92.012132} {\bibfield  {journal} {\bibinfo  {journal} {Phys. Rev. E}\ }\textbf {\bibinfo {volume} {92}},\ \bibinfo {pages} {012132} (\bibinfo {year} {2015})}\BibitemShut {NoStop}%
\bibitem [{\citenamefont {Garrahan}(2016)}]{garrahan-classical-2016}%
  \BibitemOpen
  \bibfield  {author} {\bibinfo {author} {\bibfnamefont {J.~P.}\ \bibnamefont {Garrahan}},\ }\href {https://iopscience.iop.org/article/10.1088/1742-5468/2016/07/073208} {\bibfield  {journal} {\bibinfo  {journal} {J. Stat. Mech. Theory Exp.}\ }\textbf {\bibinfo {volume} {2016}},\ \bibinfo {pages} {073208} (\bibinfo {year} {2016})}\BibitemShut {NoStop}%
\bibitem [{\citenamefont {Tang}\ \emph {et~al.}(2020)\citenamefont {Tang}, \citenamefont {Tu},\ and\ \citenamefont {Wang}}]{tang-continuous-2020}%
  \BibitemOpen
  \bibfield  {author} {\bibinfo {author} {\bibfnamefont {W.}~\bibnamefont {Tang}}, \bibinfo {author} {\bibfnamefont {H.-H.}\ \bibnamefont {Tu}},\ and\ \bibinfo {author} {\bibfnamefont {L.}~\bibnamefont {Wang}},\ }\href {https://doi.org/10.1103/PhysRevLett.125.170604} {\bibfield  {journal} {\bibinfo  {journal} {Phys. Rev. Lett.}\ }\textbf {\bibinfo {volume} {125}},\ \bibinfo {pages} {170604} (\bibinfo {year} {2020})}\BibitemShut {NoStop}%
\bibitem [{\citenamefont {Tang}\ \emph {et~al.}(2021)\citenamefont {Tang}, \citenamefont {Xie}, \citenamefont {Wang},\ and\ \citenamefont {Tu}}]{tang-tensor-2021}%
  \BibitemOpen
  \bibfield  {author} {\bibinfo {author} {\bibfnamefont {W.}~\bibnamefont {Tang}}, \bibinfo {author} {\bibfnamefont {X.~C.}\ \bibnamefont {Xie}}, \bibinfo {author} {\bibfnamefont {L.}~\bibnamefont {Wang}},\ and\ \bibinfo {author} {\bibfnamefont {H.-H.}\ \bibnamefont {Tu}},\ }\href {https://doi.org/10.1103/PhysRevD.104.114513} {\bibfield  {journal} {\bibinfo  {journal} {Phys. Rev. D}\ }\textbf {\bibinfo {volume} {104}},\ \bibinfo {pages} {114513} (\bibinfo {year} {2021})}\BibitemShut {NoStop}%
\bibitem [{\citenamefont {Tilloy}(2021)}]{tilloy-relativistic-2021}%
  \BibitemOpen
  \bibfield  {author} {\bibinfo {author} {\bibfnamefont {A.}~\bibnamefont {Tilloy}},\ }\href {https://doi.org/10.1103/PhysRevD.104.096007} {\bibfield  {journal} {\bibinfo  {journal} {Phys. Rev. D}\ }\textbf {\bibinfo {volume} {104}},\ \bibinfo {pages} {096007} (\bibinfo {year} {2021})}\BibitemShut {NoStop}%
\bibitem [{\citenamefont {Tilloy}\ and\ \citenamefont {Cirac}(2019)}]{tilloy-continuous-2019}%
  \BibitemOpen
  \bibfield  {author} {\bibinfo {author} {\bibfnamefont {A.}~\bibnamefont {Tilloy}}\ and\ \bibinfo {author} {\bibfnamefont {J.~I.}\ \bibnamefont {Cirac}},\ }\href {https://doi.org/10.1103/PhysRevX.9.021040} {\bibfield  {journal} {\bibinfo  {journal} {Phys. Rev. X}\ }\textbf {\bibinfo {volume} {9}},\ \bibinfo {pages} {021040} (\bibinfo {year} {2019})}\BibitemShut {NoStop}%
\bibitem [{\citenamefont {Shachar}\ and\ \citenamefont {Zohar}(2022)}]{shachar-approximating-2022}%
  \BibitemOpen
  \bibfield  {author} {\bibinfo {author} {\bibfnamefont {T.}~\bibnamefont {Shachar}}\ and\ \bibinfo {author} {\bibfnamefont {E.}~\bibnamefont {Zohar}},\ }\href {https://doi.org/10.1103/PhysRevD.105.045016} {\bibfield  {journal} {\bibinfo  {journal} {Phys. Rev. D}\ }\textbf {\bibinfo {volume} {105}},\ \bibinfo {pages} {045016} (\bibinfo {year} {2022})}\BibitemShut {NoStop}%
\bibitem [{\citenamefont {Hauru}\ \emph {et~al.}(2021)\citenamefont {Hauru}, \citenamefont {Van~Damme},\ and\ \citenamefont {Haegeman}}]{hauru-riemannian-2021}%
  \BibitemOpen
  \bibfield  {author} {\bibinfo {author} {\bibfnamefont {M.}~\bibnamefont {Hauru}}, \bibinfo {author} {\bibfnamefont {M.}~\bibnamefont {Van~Damme}},\ and\ \bibinfo {author} {\bibfnamefont {J.}~\bibnamefont {Haegeman}},\ }\href {https://scipost.org/10.21468/SciPostPhys.10.2.040} {\bibfield  {journal} {\bibinfo  {journal} {SciPost Phys.}\ }\textbf {\bibinfo {volume} {10}},\ \bibinfo {pages} {040} (\bibinfo {year} {2021})}\BibitemShut {NoStop}%
\bibitem [{\citenamefont {Rommer}\ and\ \citenamefont {\"Ostlund}(1997)}]{rommer-class-1997}%
  \BibitemOpen
  \bibfield  {author} {\bibinfo {author} {\bibfnamefont {S.}~\bibnamefont {Rommer}}\ and\ \bibinfo {author} {\bibfnamefont {S.}~\bibnamefont {\"Ostlund}},\ }\href {https://doi.org/10.1103/PhysRevB.55.2164} {\bibfield  {journal} {\bibinfo  {journal} {Phys. Rev. B}\ }\textbf {\bibinfo {volume} {55}},\ \bibinfo {pages} {2164} (\bibinfo {year} {1997})}\BibitemShut {NoStop}%
\bibitem [{Note1()}]{Note1}%
  \BibitemOpen
  \bibinfo {note} {As discussed in the Sec.~\ref {sec:liebliniger}, the low-energy excitations of interest in the Lieb-Liniger model are particle-hole excitations, which are classified as ``two-particle'' process in the current context, as it consists of two steps: creating a hole and then generating a particle.}\BibitemShut {Stop}%
\bibitem [{\citenamefont {Calogero}(1969)}]{calogero-ground-1969}%
  \BibitemOpen
  \bibfield  {author} {\bibinfo {author} {\bibfnamefont {F.}~\bibnamefont {Calogero}},\ }\href {https://aip.scitation.org/doi/10.1063/1.1664821} {\bibfield  {journal} {\bibinfo  {journal} {J. Math. Phys.}\ }\textbf {\bibinfo {volume} {10}},\ \bibinfo {pages} {2197} (\bibinfo {year} {1969})}\BibitemShut {NoStop}%
\bibitem [{\citenamefont {Sutherland}(1971{\natexlab{a}})}]{sutherland-quantum-1971-a}%
  \BibitemOpen
  \bibfield  {author} {\bibinfo {author} {\bibfnamefont {B.}~\bibnamefont {Sutherland}},\ }\href {https://aip.scitation.org/doi/abs/10.1063/1.1665584} {\bibfield  {journal} {\bibinfo  {journal} {J. Math. Phys.}\ }\textbf {\bibinfo {volume} {12}},\ \bibinfo {pages} {246} (\bibinfo {year} {1971}{\natexlab{a}})}\BibitemShut {NoStop}%
\bibitem [{\citenamefont {Sutherland}(1971{\natexlab{b}})}]{sutherland-quantum-1971-b}%
  \BibitemOpen
  \bibfield  {author} {\bibinfo {author} {\bibfnamefont {B.}~\bibnamefont {Sutherland}},\ }\href {https://aip.scitation.org/doi/abs/10.1063/1.1665585} {\bibfield  {journal} {\bibinfo  {journal} {J. Math. Phys.}\ }\textbf {\bibinfo {volume} {12}},\ \bibinfo {pages} {251} (\bibinfo {year} {1971}{\natexlab{b}})}\BibitemShut {NoStop}%
\bibitem [{\citenamefont {Haldane}(1988)}]{haldane-exact-1988}%
  \BibitemOpen
  \bibfield  {author} {\bibinfo {author} {\bibfnamefont {F.~D.~M.}\ \bibnamefont {Haldane}},\ }\href {https://doi.org/10.1103/PhysRevLett.60.635} {\bibfield  {journal} {\bibinfo  {journal} {Phys. Rev. Lett.}\ }\textbf {\bibinfo {volume} {60}},\ \bibinfo {pages} {635} (\bibinfo {year} {1988})}\BibitemShut {NoStop}%
\bibitem [{\citenamefont {Shastry}(1988)}]{shastry-exact-1988}%
  \BibitemOpen
  \bibfield  {author} {\bibinfo {author} {\bibfnamefont {B.~S.}\ \bibnamefont {Shastry}},\ }\href {https://doi.org/10.1103/PhysRevLett.60.639} {\bibfield  {journal} {\bibinfo  {journal} {Phys. Rev. Lett.}\ }\textbf {\bibinfo {volume} {60}},\ \bibinfo {pages} {639} (\bibinfo {year} {1988})}\BibitemShut {NoStop}%
\bibitem [{\citenamefont {Maruyama}\ and\ \citenamefont {Katsura}(2010)}]{maruyama-continuous-2010}%
  \BibitemOpen
  \bibfield  {author} {\bibinfo {author} {\bibfnamefont {I.}~\bibnamefont {Maruyama}}\ and\ \bibinfo {author} {\bibfnamefont {H.}~\bibnamefont {Katsura}},\ }\href {https://journals.jps.jp/doi/abs/10.1143/JPSJ.79.073002} {\bibfield  {journal} {\bibinfo  {journal} {J. Phys. Soc. Japan}\ }\textbf {\bibinfo {volume} {79}},\ \bibinfo {pages} {073002} (\bibinfo {year} {2010})}\BibitemShut {NoStop}%
\bibitem [{Note2()}]{Note2}%
  \BibitemOpen
  \bibinfo {note} {See \protect \url {https://github.com/tangwei94/LiebLinigerBA.jl} and \protect \url {https://github.com/tangwei94/CircularCMPS.jl} for implementations for the Bethe ansatz solution and the cMPS calculations. The scripts and raw data for the results shown in the paper can be found at \protect \url {https://github.com/tangwei94/LiebLinigerKacMoody}.}\BibitemShut {Stop}%
\bibitem [{\citenamefont {Press}\ \emph {et~al.}(2007)\citenamefont {Press}, \citenamefont {Teukolsky}, \citenamefont {Vettering},\ and\ \citenamefont {Flannery}}]{press-numerical-book-2007}%
  \BibitemOpen
  \bibfield  {author} {\bibinfo {author} {\bibfnamefont {W.~H.}\ \bibnamefont {Press}}, \bibinfo {author} {\bibfnamefont {S.~A.}\ \bibnamefont {Teukolsky}}, \bibinfo {author} {\bibfnamefont {W.~T.}\ \bibnamefont {Vettering}},\ and\ \bibinfo {author} {\bibfnamefont {B.~P.}\ \bibnamefont {Flannery}},\ }\href@noop {} {\emph {\bibinfo {title} {Numerical Recipes 3rd Edition: The Art of Scientific Computing}}}\ (\bibinfo  {publisher} {Cambridge University Press},\ \bibinfo {year} {2007})\BibitemShut {NoStop}%
\bibitem [{\citenamefont {Liao}\ \emph {et~al.}(2019)\citenamefont {Liao}, \citenamefont {Liu}, \citenamefont {Wang},\ and\ \citenamefont {Xiang}}]{liao-differentiable-2019}%
  \BibitemOpen
  \bibfield  {author} {\bibinfo {author} {\bibfnamefont {H.-J.}\ \bibnamefont {Liao}}, \bibinfo {author} {\bibfnamefont {J.-G.}\ \bibnamefont {Liu}}, \bibinfo {author} {\bibfnamefont {L.}~\bibnamefont {Wang}},\ and\ \bibinfo {author} {\bibfnamefont {T.}~\bibnamefont {Xiang}},\ }\href {https://journals.aps.org/prx/abstract/10.1103/PhysRevX.9.031041} {\bibfield  {journal} {\bibinfo  {journal} {Phys. Rev. X}\ }\textbf {\bibinfo {volume} {9}},\ \bibinfo {pages} {031041} (\bibinfo {year} {2019})}\BibitemShut {NoStop}%
\bibitem [{Note3()}]{Note3}%
  \BibitemOpen
  \bibinfo {note} {For other examples, see e.g., Refs.~\cite {hauru-riemannian-2021,tilloy-relativistic-2021}}\BibitemShut {NoStop}%
\end{thebibliography}%

\bibliographystyle{apsrev4-2}

\end{document}